\documentclass[aps,preprint,nofootinbib,showpacs]{revtex4}
\usepackage{graphicx,color,amsmath,epsfig,feynarts}
\usepackage{float}
\newcommand{\be}{\begin{equation}}
\newcommand{\ee}{\end{equation}}
\newcommand{\ba}{\begin{eqnarray}}
\newcommand{\beq}{\begin{equation}}
\newcommand{\eeq}{\end{equation}}
\newcommand{\ea}{\end{eqnarray}}
\def\bea{\begin{eqnarray}}
\def\eea{\end{eqnarray}}
\newcommand{\np}{Nucl.\,Phys.\,}

\newcommand{\pr}{Phys.\,Rev.\,}

\newcommand{\Stop}{ {\widetilde{t}} }
\newcommand{\Sbot}{ {\widetilde{b}} }

\newcommand{\wt}{\widetilde}
\def\delr            {\!\stackrel{\leftrightarrow}{\partial^\mu}\!}

\def\err#1#2{\lower2pt\hbox{ $\stackrel{\scriptstyle +#1}{\scriptstyle -#2}$}}
\def\ga{\mathrel{\raise.3ex\hbox{$>$\kern-.75em\lower1ex\hbox{$\sim$}}}}
\def\la{\mathrel{\raise.3ex\hbox{$<$\kern-.75em\lower1ex\hbox{$\sim$}}}}

\begin{document}
\preprint{%
\vbox{%
\hbox{November 2009}
}}
\title{ Higgs boson enhancement effects on squark-pair production at the LHC}

\author{Abdesslam Arhrib$^{1,2}$, Rachid Benbrik$^{3,4}$,
Kingman Cheung$^{5,6,7}$ and Tzu-Chiang Yuan$^{8}$}
\affiliation{$^1$ D\'epartement de Math\'ematiques, Facult\'e des Sciences et 
Techniques, B.P 416 Tangier, Morocco \\
 $^2$ Department of Physics, National Taiwan University,
Taipei, Taiwan \\
    $^3$ Department of Physics, National Cheng Kung University, 
Tainan 701, Taiwan \\
   $^4$ National Center for Theoretical Physics, Tainan 701, Taiwan \\
$^5$Division of Quantum Phases \& Devices, School of Physics, 
Konkuk University, Seoul 143-701, Korea \\
$^6$Department of Physics, National Tsing Hua University, Hsinchu, Taiwan \\
$^7$Physics Division, National Center for Theoretical Sciences,
Hsinchu, Taiwan \\
$^8$Institute of Physics, Academia Sinica, Nankang, Taipei 11529, Taiwan
}
\date{\today}
\begin{abstract}
  We study the Higgs boson effects on third-generation squark-pair
  production in proton-proton collision at the CERN Large Hadron
  Collider (LHC), including $\Stop \Stop^*$, $\Stop\Sbot^*$, and $\Sbot
  \Sbot^*$.  We found that substantial enhancement can be obtained
  through $s$-channel exchanges of Higgs bosons at large $\tan\beta$,
  at which the enhancement mainly comes from $b\bar b$, $b\bar c$, and
  $c\bar b$ initial states.  We compute the complete set of
  electroweak (EW) contributions to all production channels. 
{This completes previous computations in the literature}. 
We found
  that the EW contributions can be significant and can reach up to
  25\% in more general scenarios and at the resonance of the heavy
  Higgs boson. The size of Higgs enhancement is comparable or even higher
than the PDF uncertainties and {so} must be included in any reliable 
analysis. {A full analytical computation of all the EW contributions is presented}.
\end{abstract}
\pacs{12.60.Jv, 14.80.Da, 14.80.Ly}
\maketitle

\section{Introduction}
Supersymmetry (SUSY) is one of the most promising extensions of the 
standard model (SM).
Not only does it provide a natural solution to 
the gauge hierarchy problem, but also gives 
a dynamical mechanism for electroweak symmetry breaking and
a natural candidate for the dark matter.
The simplest and most popular realization of supersymmetry 
is the  minimal supersymmetric standard model (MSSM)
\cite{Haber:1984rc,Gunion-Haber,susy}.
The MSSM predicts the  existence of scalar partners to all known quarks and
leptons, via which the electroweak scale is stabilized.  
Since none of these SUSY partners have been found, the SUSY must be broken
in our present world.
Naively, we expect the SUSY particles to be heavier than their SM counter
parts; however, naturalness arguments suggest that the scale of SUSY breaking, 
and hence the masses of the SUSY particles should not exceed 
${\cal O}(1 {\rm TeV})$.  

There exist some lower mass limits on these scalar SUSY partners.
One of them comes from the direct search at the Tevatron.  Current
lower limits on the first- and second- 
squark masses are $200-300$ GeV, depending on 
the gluino mass and the neutralino mass.  
An indirect limit comes from the Higgs mass bound of 
$114.4$ GeV \cite{lep}.  It is well-known
in the MSSM that radiative corrections can lift the Higgs mass at 
its tree-level bound ($m_Z$) to the current mass bound or more 
\cite{stop-loop}.  The major correction comes from the top-stop loop. 
The current Higgs mass bound demands either (i) the top squark mass to
be of order 1 TeV, or (ii) the mixing between the left- and right-handed
top squark to be strong.
While the first condition makes the search at the LHC experiments very
difficult, the second option becomes very interesting.  Not only {
can it}
satisfy the Higgs mass bound, but also allow a relatively light top squark
as light as 200 GeV, which can certainly be produced at the LHC or may be 
even at the Tevatron.  The left-right mixing effect in the first two generations
of squark is negligible.
Furthermore, the third generation of scalar
fermions, $\wt{t}, \wt{b}$, and $\wt{\tau}$ are expected to be lighter
than the {corresponding}
scalar fermions of the first and second generations
in Grand Unified SUSY models, because
of the large Yukawa-coupling evolution.  Therefore, potentially the
top and bottom squarks are among the first SUSY particles to be
discovered at the LHC.  In this work, we focus on the third generation
squark-pair production.

There have been many works on hadronic production of top and bottom squarks
\cite{DESY}.  
The leading order (LO) cross section for diagonal top-squark pair
production $\tilde{t}_i \tilde{t}_i$ (i=1,2) via gg
scattering in hadron collisions was first calculated in Ref.~\cite{Kane:1982}.
  However, these scattering reactions only lead to
squark and antisquark pairs of the same flavor and same mass eigenstates,
i.e., $\tilde{t}_1 \tilde{t}_1^*$, $\tilde{t}_2 \tilde{t}_2^*$,
not only because gluons do not couple to $\tilde{t}_L \tilde{t}_R^*$ pair,
but also because the coupling strengths to $LL$ and $RR$ are the same.
Furthermore, such squark-pair production via 
QCD {is weakly dependent} of $\tan\beta$.  
The next-to-leading order (NLO) calculations have been improved by 
a number of authors \cite{Beenakker:1997ut,Berger:2000mp}, 
including SUSY-QCD corrections.
Recently, NLO electroweak contribution of 
$\mathcal{O}(\alpha^2_s \alpha)$ as well as the lowest-order 
$\mathcal{O}(\alpha_s\alpha + \alpha^2)$ electroweak terms
to the production of diagonal squarks were studied and shown to be sizable 
\cite{Hollik:2008yi}. It was shown in Ref~.\cite{Bornhauser:2007bf} that
the tree level electroweak contributions to the production of 
squark pairs at hadron colliders, which includes 
s-channel gauge boson exchanges as well as t and/or u channel gaugino 
exchanges, are comparable to the dominant QCD contributions in some cases. 
Note that in that study \cite{Bornhauser:2007bf} the $b\bar{b}$ 
initial-state contribution was not included.

Non-diagonal squark-pair production, like $\Stop_1 \Stop_2^*$ and 
$\Sbot_1 \Sbot_2^*$, is possible at tree level via $Z$ boson exchange, as
the $LL$ and $RR$ couplings are different \cite{mixed}.
Mixed top and bottom squark production $\Stop_i \Sbot_j^*$ is also possible 
via an intermediate $W$ boson in the $2\to 2$ subprocess
or via $gW$ fusion in the $2\to 3$ subprocess \cite{mixed}.  
All these non-diagonal $\Stop_i \Stop_j^*$, $\Stop_i\Sbot_j^*$,
and $\Sbot_i \Sbot_j^*$ pairs with $i \!\not = j$ are produced via
exchanges of electroweak gauge bosons, and therefore the 
production rates are smaller than the corresponding 
diagonal pairs.

In this work, we point out that the $s$-channel
exchange of various Higgs bosons plays an important role here, namely,
it can substantially enhance the production rate, especially at large
$\tan\beta$. 
In this paper, we investigate the importance of electroweak channels
including the neutral Higgs bosons for 
$\Stop_i \Stop_j^*$ and $\Sbot_i \Sbot_j^*$ pair production 
and charged Higgs boson for $\Stop_i\Sbot_j^*$ pair
at hadron colliders when one or both of the initial-state partons 
are the bottom quark. 
We anticipate the contributions from the Higgs exchanges will be
substantial at large $\tan\beta$ region, where the smallness of
the bottom-parton luminosity can be compensated by enhancement of the 
bottom Yukawa coupling.

Furthermore, it also allows resonant Higgs production for 
relatively heavy Higgs bosons $H^\pm$, $H^0$, and $A^0$. In fact, with
$m_{H^0, A^0, H^\pm} > m_{\tilde{q}_i} + m_{\tilde{q}_j} $, the nondiagonal
squark pair production offers an interesting possibility to study the
squark-squark-Higgs couplings right at the Higgs boson resonances.
In addition, with the intermediate charged Higgs boson all left-
and right-handed squark pairs $\tilde{t}_{L,R} \tilde{b}_{L,R}$ can be
produced while only $\tilde{t}_{L} \tilde{b}_{L}$ can be produced
via the intermediate $W$ exchange.  Thus, we anticipate the
production via intermediate charged Higgs boson could be dominant
in some region of parameter space.  We explore the MSSM parameter space
relevant for our study.

The organization is as follows.  In the next section, we will write
the details of the couplings and mass matrices.  
In Sec. III, {we describe the formulas for squark-pair 
production.
In Sec. IV, we discuss the effect of phenomenological constraints on SUSY
parameters, 
followed by numerical results on production cross sections.}
We conclude in Sec. V.

\section{Formalism}

We start with the following superpotential
\begin{equation}
W = \epsilon_{ab} \left [ y^u_{ij} Q^a_j H_u^b U^c_i 
  - y^d_{ij} Q^a_j H_d^b D^c_i 
  -  y^l_{ij} L^a_j H_d^b E^c_i 
  +  \mu  H_u^a H_d^b \right ] \;,
\end{equation}
where $\epsilon_{12}= - \,\epsilon_{21} =1$, $i,j$ are family indices,
and $y^u$ and $y^d$ represent the Yukawa matrices for the 
up-type and down-type quarks, respectively.
Here $Q, L, U^c, D^c, E^c, H_u$, and $H_d$ denote the quark doublet, 
lepton doublet,
up-type quark singlet, down-type quark singlet, lepton singlet,
up-type Higgs doublet, and down-type Higgs doublet superfields, 
respectively.

\subsection{Quark mass matrices}
The Higgs doublets develop vacuum expectation values (VEV), which
break the electroweak symmetry, are 
\begin{equation}
\langle H_u \rangle = \frac{1}{\sqrt{2}} \left(\begin{array}{c}
                                        0 \\
                                        v \sin\beta \end{array}\right )\;,
\qquad {\rm and} \qquad
\langle H_d \rangle = \frac{1}{\sqrt{2}} \left(\begin{array}{c}
                                        v \cos \beta  \\
                                        0 \end{array}\right ) \;,
\end{equation}
where $v \approx 246$ GeV, and the quark mass terms are given by
\begin{equation}
 {\cal L}_{\rm quark} =
 - y^u_{ij} \frac{v \sin\beta}{\sqrt{2}} \overline {u_{R}}_i {u_L}_j 
 - y^d_{ij} \frac{v \cos\beta}{\sqrt{2}} \overline {d_{R}}_i {d_L}_j 
  + {\rm h.c}
\;.
\end{equation}
The Yukawa matrices $y^u$ and $y^d$ are diagonalized by bi-unitary 
transformations:
\begin{equation}
\left( \begin{array}{c}
          d_1 \\
          d_2 \\
          d_3  \end{array} \right )_{L,R} = D_{L,R} \;
\left( \begin{array}{c}
          d\\
          s \\
          b  \end{array} \right )_{L,R} \,, \qquad
\left( \begin{array}{c}
          {u}_1 \\
          {u}_2 \\
          {u}_3  \end{array} \right )_{L,R} = U_{L,R} \;
\left( \begin{array}{c}
          {u} \\
          {c} \\
          {t}  \end{array} \right )_{L,R} \,,
\end{equation}
and
\begin{equation}
  U_R^\dagger \left( y^u \frac{v \sin\beta}{\sqrt{2}} \right ) U_L 
= \left( \begin{array}{ccc} 
          m_u & 0 & 0 \\
          0 & m_c & 0 \\
          0 & 0 & m_t  \end{array} \right ) \;, \qquad 
  D_R^\dagger \left( y^d \frac{v \cos\beta}{\sqrt{2}} \right ) D_L 
= \left( \begin{array}{ccc} 
          m_d & 0 & 0 \\
          0 & m_s & 0 \\
          0 & 0 & m_b  \end{array} \right ) \;.
\end{equation}
We define the diagonal Yukawa matrices $Y^u$ and $Y^d$ as
\begin{equation}
Y^u = \frac{\sqrt{2}}{ v \sin\beta } \,
  \left( \begin{array}{ccc} 
          m_u & 0 & 0 \\
          0 & m_c & 0 \\
          0 & 0 & m_t  \end{array} \right ) \;, \qquad 
Y^d = \frac{\sqrt{2}}{ v \cos\beta } \,
  \left( \begin{array}{ccc} 
          m_d & 0 & 0 \\
          0 & m_s & 0 \\
          0 & 0 & m_b  \end{array} \right ) \;.
\end{equation}

In order to avoid excessive flavor-changing neutral currents, we assume that
the squark mass matrices are in alignment with the quark mass matrices, i.e.,
they are diagonalized by the same bi-unitary transformations:
\begin{equation}
\left( \begin{array}{c}
          \tilde{u}_1 \\
          \tilde{u}_2 \\
          \tilde{u}_3  \end{array} \right )_{L,R} = U_{L,R} \;
\left( \begin{array}{c}
          \tilde{u} \\
          \tilde{c} \\
          \tilde{t}  \end{array} \right )_{L,R} \,, \qquad
\left( \begin{array}{c}
          \tilde{d}_1 \\
          \tilde{d}_2 \\
          \tilde{d}_3  \end{array} \right )_{L,R} = D_{L,R} \;
\left( \begin{array}{c}
          \tilde{d} \\
          \tilde{s} \\
          \tilde{b}  \end{array} \right )_{L,R} \,,
\end{equation}
Here 
$d_1,d_2,d_3$ are in interaction basis while $d,s,b$ are in mass eigenbasis.
Without loss of generality, we can make the choice that 
the right-handed quarks and squarks
are already in the mass eigenbasis, ie., $D_R = U_R = I$. The information
on the left-handed unitary matrices $D_L, U_L$ is encoded in the 
Kobayashi-Maskawa (KM) matrix as
\begin{equation}
   V_{KM} = U^\dagger_L \, D_L \;.
\end{equation}

\subsection{Squark mass matrices}

Even after we have rotated the quark and squark mass matrices into 
family-diagonal form with the same bi-unitary transformation, within
one family the left-handed and right-handed squarks will mix due to the
soft terms and $F$ terms in the Lagrangian.  This $LR$ mixing is
proportional to the quark mass concerned such that the $LR$ mixing for the
first two generations are negligible while it can be substantial for
the third generation.  From now on we only concern the $LR$ squark 
mixing in the third generation.

The squark mass-squared matrix in the $L$-$R$ basis have the form 
\begin{equation}
{\cal M}^2_{\tilde{q}}= 
     \left( \begin{array}{cc} 
                m_{LL}^2 & m^2_{LR}  \\
                m^{2\star}_{LR}     &  m_{RR}^2
            \end{array} \right) \label{M-stop}
\end{equation}
with
\begin{eqnarray}
  m_{LL}^2 &=& \widetilde{M}_{L}^2 + m_q^2
    + m_Z^2\cos 2\beta\,( I_3^q - e_q s_W^2 ) ,\\
  m_{RR}^2 & = & \widetilde{M}_{R}^2  + m_q^2
                  + m_Z^2 \cos 2\beta\, e_q s_W^2 , \label{eq:c} \\[2mm]
  m_{LR}^2    &=& 
  m_q \left [A_q^\star - \mu\ (\tan\beta)^{-2I_3^q} \right ]\,\, , \label{eq:d}
\end{eqnarray}
where 
$I_3^q=\pm 1/2$ and $e_q$ are the third component of the 
weak isospin and the electric charge of the quark $q$.
In Eqs.~(\ref{M-stop}),  $\mu$ is the supersymmetric Higgs mass parameter,
$\wt{M}^2_{L}$ the soft-breaking mass parameter for the 
squark iso-doublet $(\wt{q}_L, \wt{q'}_L)$, and
$\wt{M}^2_{R}$  are the soft-breaking 
mass parameters for the iso-singlets $\wt{q}_R$.
They can be different for each generation, but for simplicity
we will assume equal values for all generations  
$\widetilde{M}_{L}=\widetilde{M}_{R}=M_{SUSY}$ in 
our numerical analysis.
$A_q$ are the parameters of the soft-breaking scalar three-point 
interactions of top- and bottom-squarks with the Higgs fields.

The hermitian matrix { in Eq.~(\ref{M-stop})} is 
diagonalized by  a unitarity matrix 
$R^{{\widetilde{q}}}$, which rotates the current eigenstates,
${\widetilde{q}}_L$ and ${\widetilde{q}}_R$, into the mass
eigenstates $\widetilde{q}_1$ and $\widetilde{q}_2$ as follows,
\begin{equation}
\left(\begin{array}{c} 
                {\widetilde{q}}_1 \\
                {\widetilde{q}}_2
\end{array} \right) = R^{{\widetilde{q}}} 
\left(\begin{array}{c} 
{\widetilde{q}}_L \\
{\widetilde{q}}_R
\end{array} \right)  =
\underbrace{ \left( \begin{array}{cc} 
 \cos{\wt{\theta}_q}  & 
 \sin{\wt{\theta}_q}   \\
- \sin\wt{\theta}_q &
 \cos\wt{\theta}_q
            \end{array} \right)}_{R^{{\widetilde{q}}} } 
\left( \begin{array}{c} 
                {\widetilde{q}}_L \\
                {\widetilde{q}}_R 
\end{array} \right) \, ,
\label{eqe}
\end{equation}
yielding the physical mass eigenvalues, with the convention
$m_{{\widetilde{q}}_{1}}< m_{{\widetilde{q}}_{2}}$,  
\begin{eqnarray}
m_{{\widetilde{q}}_{1,2}}^2 &=& \frac{1}{2}\left( m_{LL}^2 + m_{RR}^2 
\mp \sqrt{ (m_{LL}^2 - m_{RR}^2)^2 + 4 |m_{LR}^2|^2  }\, \right)\ . 
\label{mass} 
\end{eqnarray}
The mixing angle $\wt{\theta}_q$ obeys the relation  
\begin{eqnarray}
\tan 2\wt{\theta}_q =\frac{2 m^2_{LR} }{m_{LL}^2 -m_{RR}^2 } \ \ . 
\label{mixing}
\end{eqnarray} 
%
%
Hence, for the  case of the supersymmetric partners of the light fermions, 
$L$--$R$ mixing can be neglected.
However, mixing between top squarks can be
sizable and allows one of the two mass eigenstates
to be lighter than the top quark. Bottom-squark 
mixing can also be significant if $\tan\beta$ is large.

\subsection{Higgs and gauge bosons 
interactions with quarks and squarks}

Let $H_k=(h^0, H^0, A^0, G^0)$ ($k$=1...4), one can write 
the relevant Lagrangian density in the
($\tilde{q}_1,\tilde{q}_2$) basis as following form ($i,j$=1,2)
%
%
\begin{eqnarray}
\label{lagrangian}
{\cal L}_{\rm relevant} &=& g_{H_kqq} H_{k}\bar{q}q + 
(G_k)_{ij}H_k\tilde{q}_j^{\ast}\tilde{q}_i + 
(G_5)_{ij}H^+\tilde{q}_j^{\ast}\tilde{q}_i   \\
&+&  g\bar{q}({{\cal A}}_{im}^{\tilde{q}}P_R
+{{\cal B}}_{im}^{\tilde{q}})\tilde{\chi}^0_m\tilde{q}_i 
+g\bar{q^\prime}({{\cal L}}_{il}^{\tilde{q}}P_R
+{{\cal K}}_{il}^{\tilde{q}}P_L)\tilde{\chi}_l^{+c}\tilde{q}_i \nonumber \\
&+&H^+ \, \bar{u} \left(
    Y^u \cos\beta P_L + Y^d \sin\beta P_R \right)\, V_{KM} d  \nonumber \\
&+& G^+ \bar{u} \left(
    Y^u \sin\beta P_L - Y^d \cos\beta P_R \right)\, V_{KM} d  \nonumber \\
 &+& {\rm h.c}\nonumber
\end{eqnarray}
The Feynman rules are $ig_{H_kqq}$ for k =1,2 and $\gamma_5g_{H_kqq}$ for k= 3,4 where we list only Higgs bosons couplings to quark b, and
\begin{eqnarray}
g_{hbb} = \frac{g\, m_b}{2m_W}\frac{\sin\alpha}{\cos\beta}, \quad g_{Hbb} = -\frac{g\, m_b}{2m_W}\frac{\cos\alpha}{\cos\beta},\quad g_{Abb} = -\frac{g\, m_b}{2m_W}\tan\beta.
\end{eqnarray}
The couplings  {$G_k$ and $G_5$ are given in the mass basis by}
\begin{eqnarray}
G_k ={{\cal R}}^{\tilde{q}}\hat{G}_k ({{\cal R}}^{\tilde{q}})^T,
\hspace{0.6cm} (k=1...4),\hspace{1.0cm}
G_{5} ={{\cal R}}^{\tilde{t}}
\hat{G}_{5} ({{\cal R}}^{\tilde{b}})^T 
\end{eqnarray}

where $\hat{G}_k$ and $\hat{G}_5$  are the couplings in the
($\tilde{q}_L,\tilde{q}_R$) basis, and their explicit forms are
shown in Appendix \ref{appendix-hsc}. Finally,
${{\cal A}}_{im}^{\tilde{q}}$, ${{\cal B}}_{im}^{\tilde{q}}$
(m=1...4), and ${{\cal L}}_{il}^{\tilde{q}}$, ${{\cal K}}_{il}^{\tilde{q}}$ (l=1,2)
used in Eq.(\ref{lagrangian}) are defined also in the end of Appendix \ref{appendix-hsc}.

The $W$ and Z bosons interactions with the quarks are given by
\begin{equation}
{\cal L} = - \frac{g}{\sqrt{2}}\overline{u} \gamma^\mu V_{KM} P_L d 
   W^+_\mu - g Z \bar{q} \gamma^\mu (C_{qL}P_L + C_{qR}P_R)q + {\rm H.c.}
\end{equation}
with
\begin{equation}
  C_{\!q L,R} \,=\frac{1}{\cos\theta_W} \big(I_{3L,R}^q - 
e_q \sin^2\theta_W \big)
\end{equation}

The $W$ and $Z$ bosons interactions with the squarks are given by
\begin{equation}
 {\cal L} = - ig  g_{W\tilde{q}_i \tilde{q}_j}W^{+}_{\mu}  
    \widetilde{u}^{*}_j \delr  \widetilde{d}_i 
 - ig g_{Z\tilde{q}_i \tilde{q}_j}Z \tilde{q}^*_i \delr \tilde{q}_j 
        + {\rm h.c.}
\end{equation}
with
\begin{eqnarray}
 g_{W\tilde{q}_i \tilde{q}_j}  &=& \frac{1}{\sqrt{2}} 
\left( \begin{array}{cc}
    \cos\theta_{\tilde{u}}\cos\theta_{\tilde{d}}  
         & - \cos\theta_{\tilde{u}}\sin\theta_{\tilde{d}}\\[1mm]
    - \sin\theta_{\tilde{u}}\cos\theta_{\tilde{d}}
         & \sin\theta_{\tilde{u}}\sin\theta_{\tilde{d}}
  \end{array} \right)_{ij}
\label{eq:wij}
\end{eqnarray}
and
\begin{eqnarray}
 g_{Z\tilde{q}_i \tilde{q}_j} &=& \frac{1}{\cos\theta_W} \left( \begin{array}{cc}
    I^{q}_{3L}\,\cos^2\theta_{\tilde{q}} - e_q \sin^2\theta_W 
         & -\frac{1}{2}\, I^{q}_{3L}\,\sin 2\theta_{\tilde{q}} \\[1mm]
    -\frac{1}{2}\, I^{q}_{3L}\,\sin 2\theta_{\tilde{q}} 
         & I^{q}_{3L}\,\sin^2\theta_{\tilde{q}} - e_q \sin^2\theta_W
  \end{array} \right)_{ij}
\label{eq:cij}
\end{eqnarray}

%
\subsection{Radiative corrections to the Yukawa couplings}
Note that both initial and final states
have the  bottom Yukawa $Y^b$ dependence. 
It is now well established  that the coupling of the 
scalar bottom $\wt{b}$ to the up-type Higgs doublet
induces a modification of the tree-level relation between the 
bottom quark mass and its Yukawa coupling \cite{Hall:1993gn,Carena:1999py,Carena:1994bv,Pierce:1996zz}. 
Those corrections are amplified at large $\tan\beta$.
The  modifications can be absorbed by redefining the bottom Yukawa
coupling as 
\cite{Hall:1993gn,Carena:1999py,Carena:1994bv,Pierce:1996zz}
\begin{equation}
Y^b = \frac{\sqrt{2} m_b }{v \cos\beta} 
 \rightarrow \frac{\sqrt{2}}{v \cos\beta} \frac{m_b}{1 + \Delta_b}\, 
\approx \frac{\sqrt{2}}{v} \frac{m_b}{1 + \Delta_b} \tan\beta
\end{equation} 
where the second term is valid for large $\tan\beta$ 
and {the SUSY-QCD corrections lead to}
\begin{equation}
\Delta_b =  \frac{2\alpha_s }{3\pi} \,\mu \,m_{\wt g} 
\, \tan\beta \,I(m_{\wt b_1}, m_{\wt b_2}, m_{\wt g}) 
+ \frac{(Y^t)^2 }{16 \pi^2} \,\mu \,A_t \,\tan \beta \,
  I ( m_{\wt t_1}, m_{\wt t_2}, \mu )
\label{deltab}
\end{equation}
where $m_{\wt g}$ is the gluino mass, and
the function $I$ is given by
\begin{eqnarray}
I(a,b,c) = \frac{1}{(a^2-b^2)(b^2-c^2)(a^2-c^2)}
(a^2 b^2\ln\frac{a^2}{b^2}+b^2 c^2\ln\frac{b^2}{c^2}+
c^2 a^2\ln\frac{c^2}{a^2}) 
\label{ifun} \quad .
\end{eqnarray}
In $\Delta_b$ we only keep the dominant contributions from the gluino-sbottom
and charged-higgsino-stop loops because they are proportional to the strong
coupling and to the top Yukawa coupling, respectively, while neglecting 
those that are proportional to the weak gauge coupling.  Note that $\Delta_b$
is evaluated at the scale of SUSY particles $M_{\rm susy}$, where the heavy
particles in the loop decouple, whereas the bottom Yukawa coupling  $Y^b(Q)$ 
at any scale $Q$ is determined by the running $b$-quark mass $m_b(Q)$
at the scale $Q$:

The contributions  to the bottom Yukawa couplings
which are enhanced at large $\tan\beta$ can be included
to all orders by making 
the following 
replacements~\cite{Hall:1993gn,Carena:1999py,Guasch:2003cv,Carena:2006ai}
\begin{eqnarray}
g_{hbb} & \rightarrow & g_{hbb} \frac{1 - \Delta_b/(\tan\beta \tan\alpha)}{1 + \Delta_b} \\
g_{Hbb} & \rightarrow & g_{Hbb} \frac{1 + \Delta_b \tan\alpha /\tan\beta}{1 + \Delta_b} \\
g_{Abb} & \rightarrow & g_{Abb} \frac{1 - \Delta_b/\tan^2\beta}{1 + \Delta_b} \\
 {Y^b} & \rightarrow & 
{\frac{ \sqrt{2} m_b (Q)}{ v \cos\beta }
\frac{1}{1+\Delta_b}}
\label{yuk-cor}
\end{eqnarray}
We now have all the tools to compute the production cross section
of sbottom and stop.

\section{Squark Pair Production}
In this section we discuss squark-pair production.
Let us define our notation for the convenience of the following formulas.  
The momenta of the incoming quark $q$ and anti-quark $\bar q$, 
outgoing squark $\widetilde{q}_i$ and outgoing anti-squark $\widetilde{q}_j^*$
are denoted by $p_1$, $p_2$, $k_1$ and $k_2$, respectively.
We neglect the quark masses of the incoming partons.  
The Mandelstam variables are defined as follows
\begin{eqnarray}
\hat s &=&  (p_1+p_2)^2 = (k_1+k_2)^2  \nonumber \\
\hat t &=& (p_1-k_1)^2 = (p_2-k_2)^2 
= \frac{ m_{\wt{q}_i}^2 + m_{\wt{q}_j}^2}{2} - 
\frac{\hat s}{2} \left( 1 - \beta \cos\theta^* \right ) \nonumber \\
\hat u &=& (p_1-k_2)^2 = (p_2-k_1)^2 
= \frac{ m_{\wt{q}_i}^2 + m_{\wt{q}_j}^2}{2} - 
\frac{\hat s}{2} \left( 1 + \beta \cos\theta^* \right ) 
\label{mandelstan}
\end{eqnarray}
where $\beta = \lambda^{1/2}(1, m_{\wt{q}_i}^2/\hat s, m_{\wt{q}_j}^2/\hat s)$ 
and $\theta^*$ is
the scattering angle in the center-of-mass frame of the partons.  
\begin{figure}
\centering
\vspace{-2cm}
\input{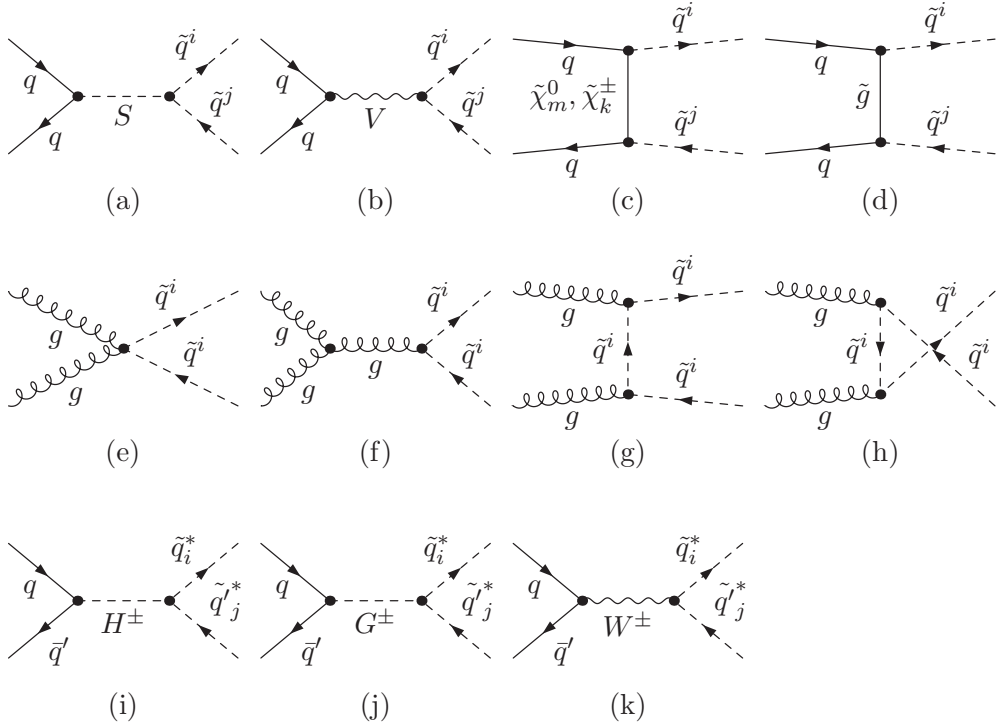}
\vspace{-7cm}
\caption{Tree-level Feynman diagrams for squarks pair production  via 
$q\bar{q}$, $gg$, and $q\bar q'$ annihilation. 
With (i,j) = (1,2), $q= (u, d, c, s, b)$, 
$S = h^0, H^0, A^0, G^0$ and $V = \gamma, Z, g$. In diagram (b),
the photon and gluon do not contribute to non-diagonal production.}
\label{fig:qqsts-diagrams}
\end{figure}
\begin{figure}[h]
\centering
\input{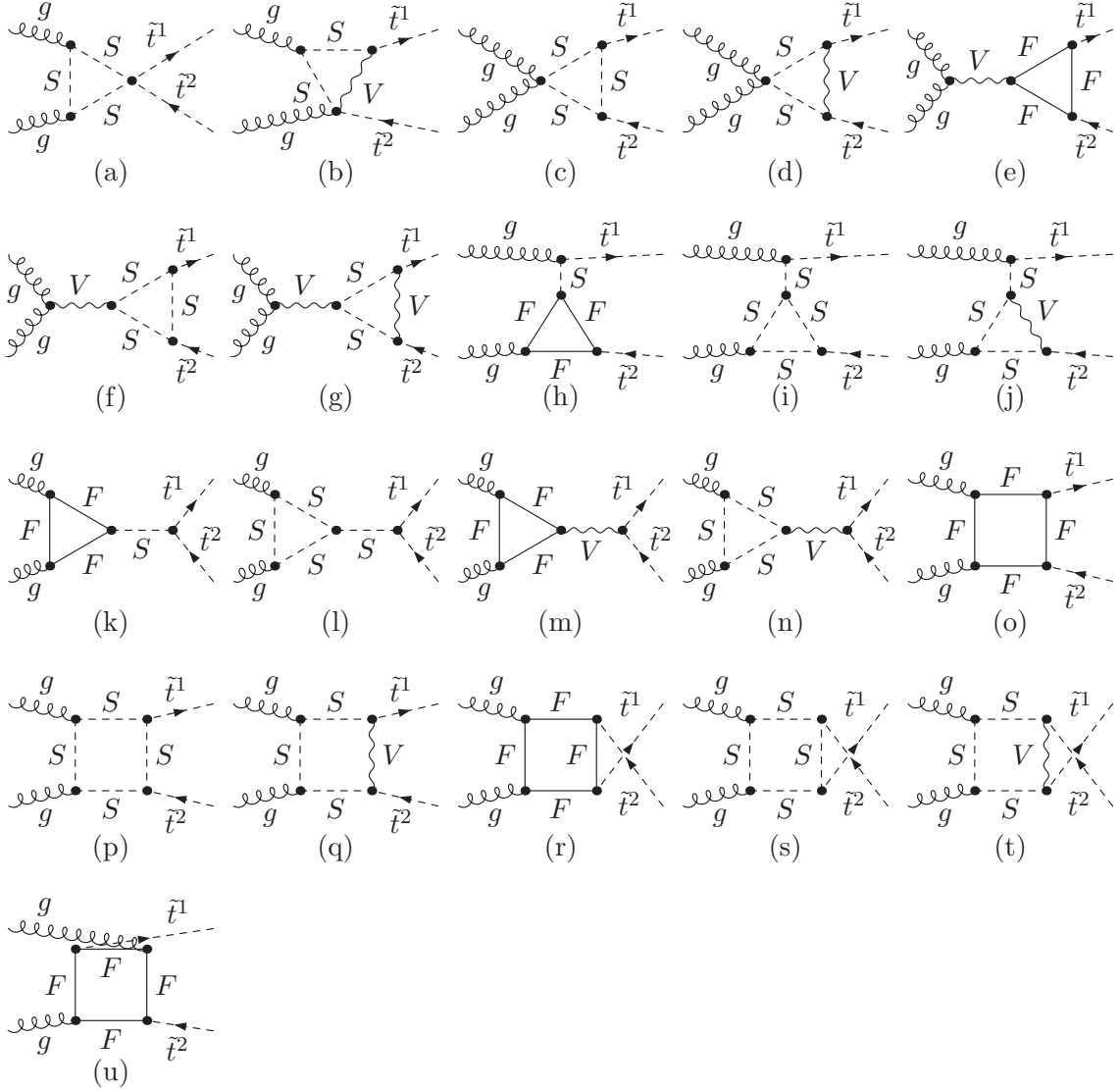}
\caption{One-loop Feynman diagrams for squark-pair production 
in gluon-gluon fusion. With i $\neq$ j = (1,2), $F= \tilde{g}, \tilde{\chi}^0,
 \tilde{\chi}^\pm$, $S = h^0, H^0, A^0, G^0$ and $V = \gamma, Z, g$.} 
\label{fig:ggsts-diagrams}
\end{figure}

\subsection{Hadronic production of $q\bar{q}, gg \to
  \tilde{t}_{i}\tilde{t}^*_{j},  \tilde{b}_{i}\tilde{b}^*_{j}$}

The production of top and bottom squark pair proceeds via the following
$q\bar{q}$- and $gg$- initiated subprocesses which are depicted in
 Fig.~\ref{fig:qqsts-diagrams}(a) -- (h).
\begin{eqnarray}
q\bar{q}, gg \to \tilde{t}_{i}\tilde{t}^*_{j}, \tilde{b}_{i}\tilde{b}^*_{j},
\end{eqnarray}
where $(i,j)=(1,2)$.

 

Note that the photon and gluon do not contribute to production of of
$\tilde{q}_1\tilde{q}_2^*$, because of electromagnetic or color
conservation.
 Consequently, at tree level the above
reactions proceed only through s-channel Z boson and Higgs bosons.
 If the initial state is $u\bar{u}$, $d\bar{d}$, $s\bar{s}$
or $c\bar{c}$, the Yukawa couplings are so small that we consider
only the contribution from the $Z$ exchange diagram. While in the case
of  $b\bar{b}$, the cross section will directly determined by the size of
Higgs coupling to a squarks  $S\tilde{q}_1\tilde{q}_2$  where $S = h^0, H^0$ or
$A^0$ and $Z\tilde{q}_1\tilde{q}_2$ couplings, which are proportional to
$\sin2\theta_{\tilde{q}}$ of the squark-mixing angle
$\theta_{\tilde{q}}$. therefore, these processes can be used to probe the
mixing angle $\theta_{\tilde{q}}$.
{The analytic} expressions for  
$b\bar b\to \tilde{t}_{i}\tilde{t}^*_{j}$
and $b\bar b\to \tilde{b}_{i}\tilde{b}^*_{j}$ are given in appendix
 \ref{cross-sections}.

The gluon-gluon fusion into $\widetilde{t}_1 \widetilde{t}_2^*$ and 
$\widetilde{b}_1 \widetilde{b}_2^*$ only goes through loop diagrams, as shown
in Fig.~\ref{fig:ggsts-diagrams}. { These one loop contributions 
are of the order of either $\alpha_s^2$ or $\alpha_s\alpha$.}

The hadronic inclusive cross section for
$\tilde{q}_i\tilde{q}_j^*$ production in proton-proton collisions at a
total hadronic center of mass energy $\sqrt{S}$ can be written as
\cite{Brock:1993sz}
\begin{eqnarray}
\sigma_{pp\to\tilde{q}_i\tilde{q}_j^*}(S) = \sum_{q} \int^1_{\tau_0} d\tau 
\frac{d{\cal{L}}^{pp}_{q\bar{q}}}{d\tau} \hat{\sigma}_{LO} 
(q\bar{q}\to \tilde{q}_i\tilde{q}_j^*)(\tau S) + \int^1_{\tau_0} d\tau 
\frac{d{\cal{L}}^{pp}_{gg}}{d\tau} \hat{\sigma}_{LO} 
(gg\to \tilde{q}_i\tilde{q}_j^*)(\tau S)
\end{eqnarray}
where $\tau_0 = (m_{\tilde{q}_i}^2 + m_{\tilde{q}_j}^2 )^2 / S$, and 
the parton luminosity is
\begin{eqnarray}
\frac{d{\cal{L}}^{pp}_{ab}}{d\tau} = \int^1_{\tau} \frac{dx}{x} 
\frac{1}{1+\delta_{ab}} \big[ f_{a}(x,\mu_F) f_{b}(\frac{\tau}{x}, \mu_F) 
+ f_{b}(x, \mu_F) f_{a}(\frac{\tau}{x}, \mu_F)\big ]  
\end{eqnarray}
where $f_{a}(x, \mu_F)$ is parton distribution functions (PDF) for
each type $a$ in the proton carrying a fraction $x$ of the proton
momentum at scale $\mu_F = m_{\tilde{q}_i} + m_{\tilde{q}_j}$.


\subsection{Hadronic production of $d\bar{u}\to \wt{t}_i \wt{b}^*_j$}
In this case, the Mandelstam variables
are defined as in Eq.~(\ref{mandelstan}) with $m_{\wt{q}_i}=m_{\wt{t}_i}$ and
$m_{\wt{q}_j}=m_{\wt{b}_j}$. 
At hadron colliders, the production { mechanism
proceeds via the conventional Drell-Yan
prcoesses with the charged Higgs boson, charged Goldstone and charged gauge 
bosons,} as depicted in Fig.~\ref{fig:qqsts-diagrams} (i)-(k).
The $s$-channel diagram with charged Higgs exchange 
dominates when $\sqrt{\hat s}$ is close to $M_{H^\pm}$.
The analytic expressions for { the cross sections}
are given in appendix  \ref{cross-sections}.

{
\section{Numerical Results}

\subsection{Phenomenological constraints}
}
In this section, and before presenting our numerical results, 
we would like to list the phenomenological constraints 
included when determining the allowed parameter space 
\cite{Djouadi:2005an,Dedes:2003cg}. 
\begin{itemize}
\item The most stringent constraint generally arises from $\Delta\rho^{SUSY}$
which receives contributions from both stop and sbottom. The extra
contributions to the $\Delta\rho^{SUSY}$ parameter from the stop and
sbottom sector \cite{Djouadi:1996pa,hagiwara} should not exceed the current
limit from precision measurements \cite{PDG}: $\Delta\rho^{SUSY}$
$\leq$ $ 10^{-3}$.

\item The soft SUSY-breaking parameters $A_q$ at the weak scale should not
be too large in order to keep the radiative corrections to the Higgs
masses under control.  In particular the trilinear couplings of the
third generation squarks $A_{t,b}$, they will play a particularly
important role in the MSSM squarks/Higgs sectors. These parameters can
be constrained in at least one way, besides the trivial requirement
that it should not make the off-diagonal term of the squark mass
matrices too large to generate too low masses for the
squarks. $A_{t,b}$ should not be too large to avoid the occurrence of
charge and color breaking (CCB) minima in the Higgs potential. To
avoid such minima at tree level, $A_{t,b}$ {have} to satisfy the
following tree level conditions\cite{Casa:1996ca}.
\begin{eqnarray}
A^2_t &\le& 3\big( m^2_{\tilde{t}_2} + m^2_{\tilde{t}_1} -2 m^2_t +\frac{1}{2} m^2_Z \cos2\beta + M^2_{H_2} + \mu^2 \big),\\
A^2_b &\le& 3\big( m^2_{\tilde{b}_2} + m^2_{\tilde{b}_1} -2 m^2_b +\frac{1}{2} m^2_Z \cos2\beta + M^2_{H_1} + \mu^2 \big).
\end{eqnarray}
With $M^2_{H_1} = (m^2_{A^0} + m^2_Z)\sin^2\beta -1/2 m^2_Z$ and 
$M^2_{H_1} + M^2_{H_2} = m^2_{A^0}$. The above constraints depend
 on $\mu$ and $A_{t,b}$ explicitely. 

\item Another constraint which has been imposed is the perturbativity of the
bottom Yukawa coupling. Since radiative corrections to the bottom Yukawa
coupling have been implemented as in Eq.~(\ref{yuk-cor}), the Yukawa may blow 
up when SUSY parameters varies. Thus, we restrict $Y^b \la (4 \pi)^2$.

\item We have imposed also all the experimental bounds on {
squark, chargino, and neutralino masses} as well as Higgs boson 
masses \cite{PDG}.

\end{itemize}

\subsection{Numerical results}
In this section, we present the numerical result for inclusive production
cross section of diagonal, non-diagonal and mixed squarks at the LHC 
with a proton-proton center-of-mass energy cross sections of 14 TeV.
In our numerical calculations the following SM input parameters were chosen 
\cite{PDG}:
\begin{eqnarray}
m_t &=& 171.9 \,\, {\rm GeV}\quad , \quad m_W = 80.398 \,\,{\rm GeV}\quad , 
\quad m_Z = 91.1878 \,\,{\rm GeV} \nonumber \\
 G_{F} &=& 1.16637 \times 10^{-5} \,\,{\rm GeV}^{-2} \quad , 
\quad V_{cb}=0.04 , \quad m_b(m_b) = 4.25 \,\,{\rm GeV}
\end{eqnarray}
The running QCD coupling $\alpha_s$ was evaluated at the two-loop level 
\cite{Gorishnii:1990zu} and the CTEQ6L PDFs \cite{Nadolsky:2008zw,Brock:1993sz} 
were used to calculate the various
cross sections. 
Moreover, in order to improve the perturbative calculations,
1-loop running masses $m_b(Q)$ were taken as following:
\begin{eqnarray}
m_b(Q) = m_b^{\rm{\overline{DR}}}(Q) = m_b^{\rm{\overline{MS}}}(Q) 
\left( 1 + \frac{4 \alpha_s}{3\pi}\right)
\end{eqnarray}
where $m_b^{\rm{\overline{MS}}}$ includes the SM QCD corrections.

{First of all,}
 we investigate the effect of varying the MSSM parameters for 
which the correction to $\Delta_b$ term are expected to have a 
large impact. 
In the limit where the squark and gluino masses 
have approximately the same value, denoted by the common SUSY mass $M_{SUSY}$,
the Eq.~(\ref{ifun}) simplifies to 
\begin{eqnarray}
  I (m_{SUSY},m_{SUSY},m_{SUSY}) = \frac{1}{2 m^2_{SUSY}}.
\end{eqnarray}
Furthermore, if $\mu$ is of similar size, 
{the first term in  Eq.~(\ref{deltab}) 
is dominant and} reduces to
\begin{eqnarray}
\Delta_b \approx {\rm sign}(\mu) \frac{\alpha_s}{3\pi} \,\,\tan\beta
\end{eqnarray}
So, for large $\tan\beta$ this effect can be ${\cal{O}}(1)$ 
and does not vanish for a heavy SUSY spectrum. 
{
The sign of $\mu$ is the 
decisive factor in determining whether 
the corrections will enhance 
or suppress the cross section for the processes of 
$pp \to b\bar{b} \to \tilde{b}_i \tilde{b}_j^*,\; 
\tilde{t}_i \tilde{t}_j^*,\; \tilde{t}_i \tilde{b}_j^*$.
}
We assume the universality of soft SUSY breaking trilinear 
couplings: $A_t = A_b = A_\tau = A_0$.
We parameterize the squark sector using the following input parameters: 
$\tan\beta$, $\mu$, $A_0$ 
and the gluino mass $m_{\tilde{g}}$. The MSSM Higgs sector is 
parameterized by the mass of CP-odd $m_{A^0}$ and $\tan\beta$ as well
as by $M_{SUSY}$, $A_{b,t}$ and $\mu$ for higher order corrections
\cite{Heinemeyer:1998yj,CPSUSYH}. All the MSSM Higgs masses and 
parameters are computed with FeynHiggs code \cite{Heinemeyer:1998yj}.

\begin{figure}[htbp] 
\begin{picture}(320,260)
\put(-80,0){\mbox{\psfig{file=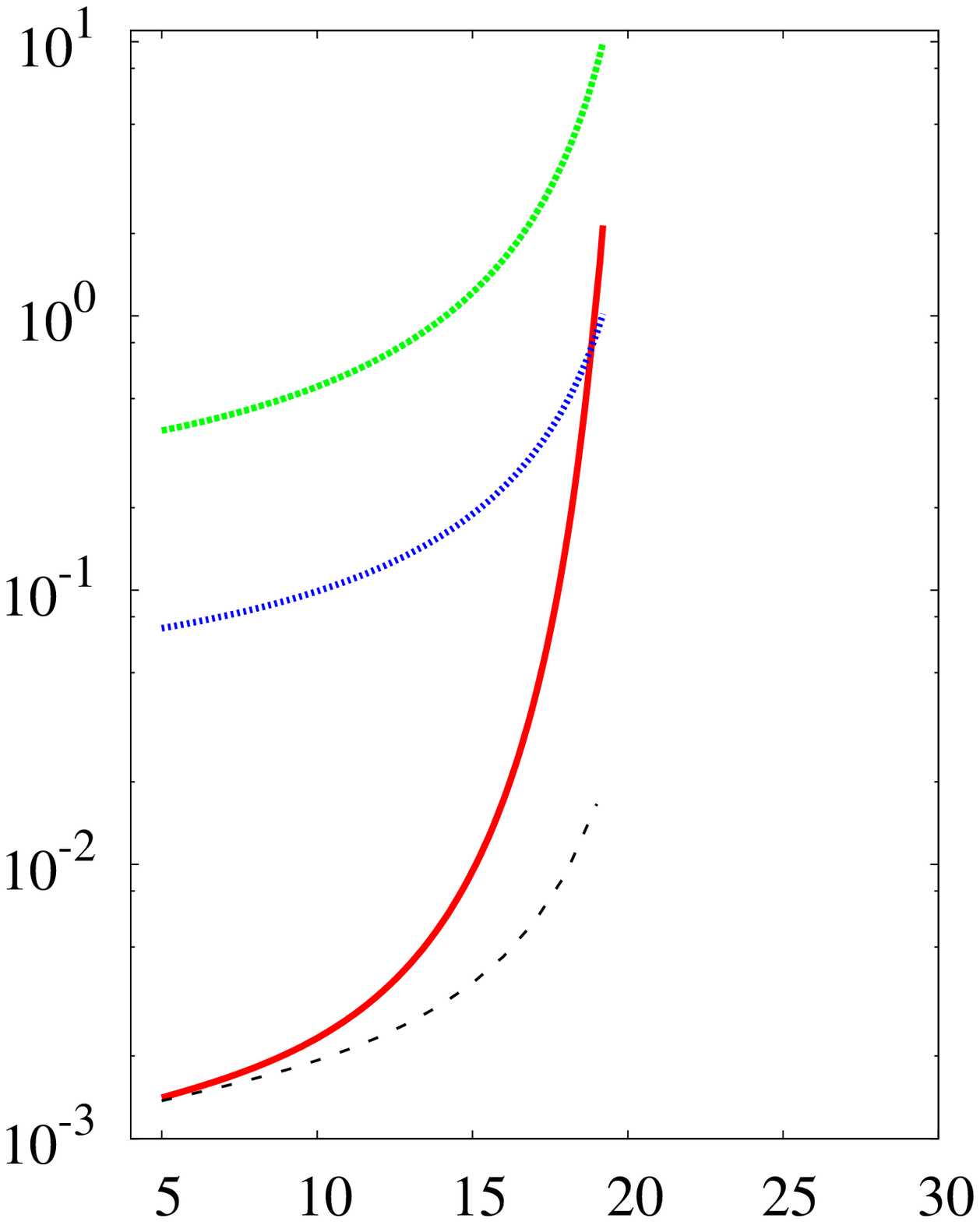,height=3.8in,width=3.2in}}}
\put(30,-10){\makebox(0,0)[bl]{\large{$\tan\beta$}}}
\put(-30,235){\makebox(0,0)[bl]{\large{$\sigma(\tilde{b_1}\tilde{b}^\star_1)$ [pb]}}}
\put(-20,195){\makebox(0,0)[bl]{\large{$gg $}}}
\put(-26,150){\makebox(0,0)[bl]{\large{$\sum q\bar{q}$}}}
\put(0,70){\makebox(0,0)[bl]{\large{$b\bar{b}$}}}
\put(30,46){\makebox(0,0)[bl]{\large{no-Higgs}}}
\put(30,30){\makebox(0,0)[bl]{\large{$m_{A^0} = 540$~GeV}}}
\put(150,0){\mbox{\psfig{file=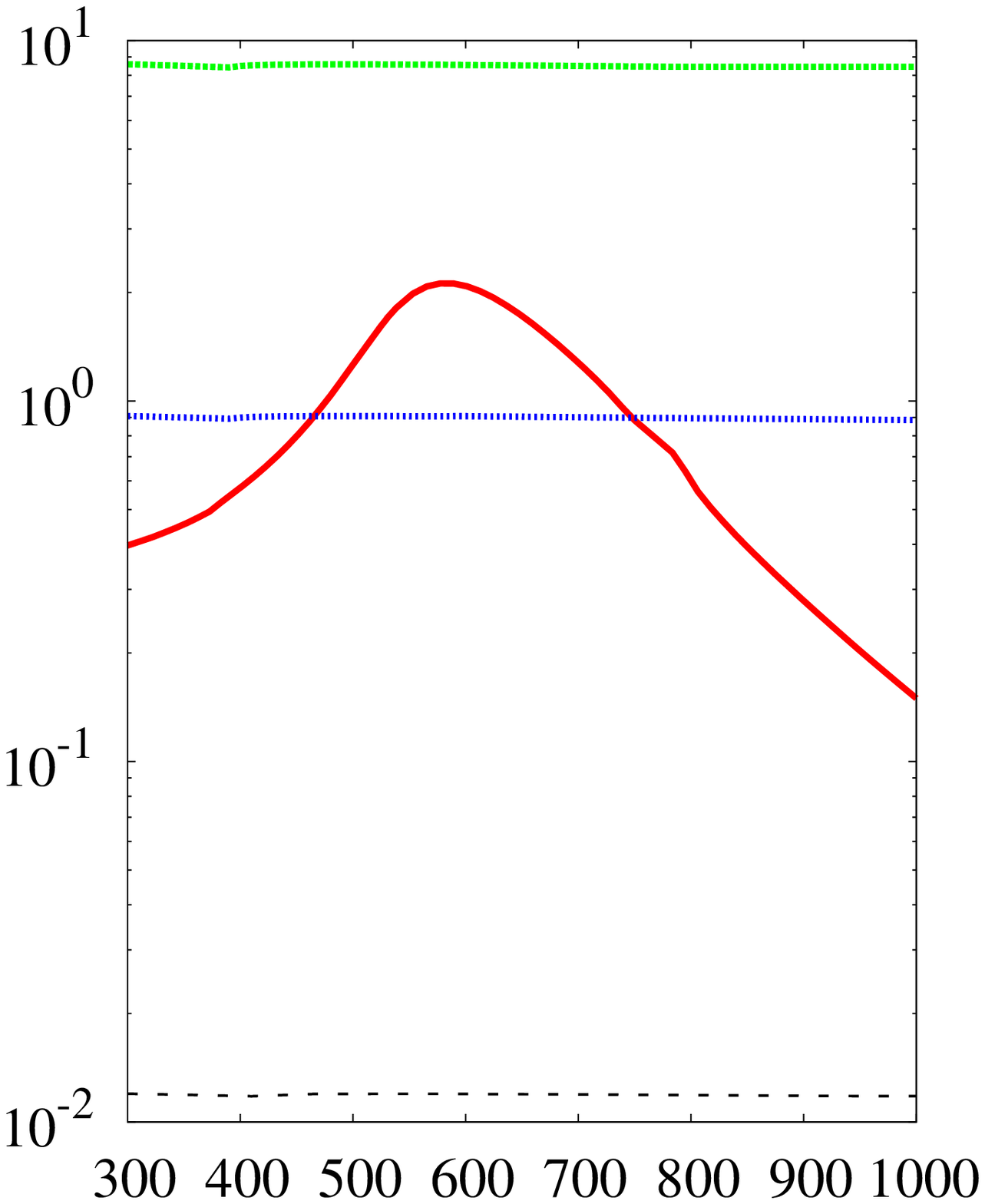,height=3.8in,width=3.2in}}}
\put(250,-10){\makebox(0,0)[bl]{\large{$m_{A^0}\,\,(GeV) $}}}
\put(250,55){\makebox(0,0)[bl]{\large{$\sigma(\tilde{b_1} \tilde{b}^\star_1)$ [pb]}}}
\put(250,240){\makebox(0,0)[bl]{\large{$gg $}}}
\put(250,190){\makebox(0,0)[bl]{\large{$b\bar{b} $}}}
\put(310,183){\makebox(0,0)[bl]{\large{$\sum q\bar{q}$}}}
\put(210,83){\makebox(0,0)[bl]{\large{$\tan\beta = 18.5$}}}
\put(210,30){\makebox(0,0)[bl]{\large{no-Higgs}}}
\end{picture}
\caption{ Production rates for $\tilde{b}_1 \tilde{b}^*_1$ pair 
production as   a function of (left) $\tan\beta$ and (right) $m_{A^0}$. 
{Other SUSY parameters are chosen to be $M_{SUSY} = 490$
  GeV,} $M_2 = 200$ GeV, $m_{\tilde{g}} = - \mu =
  1$ TeV, $ A_0 = 1140$ GeV.}
\label{fig-sb1sb1}
\end{figure}

\begin{figure}[htbp] 
\begin{picture}(320,260)
\put(-80,0){\mbox{\psfig{file=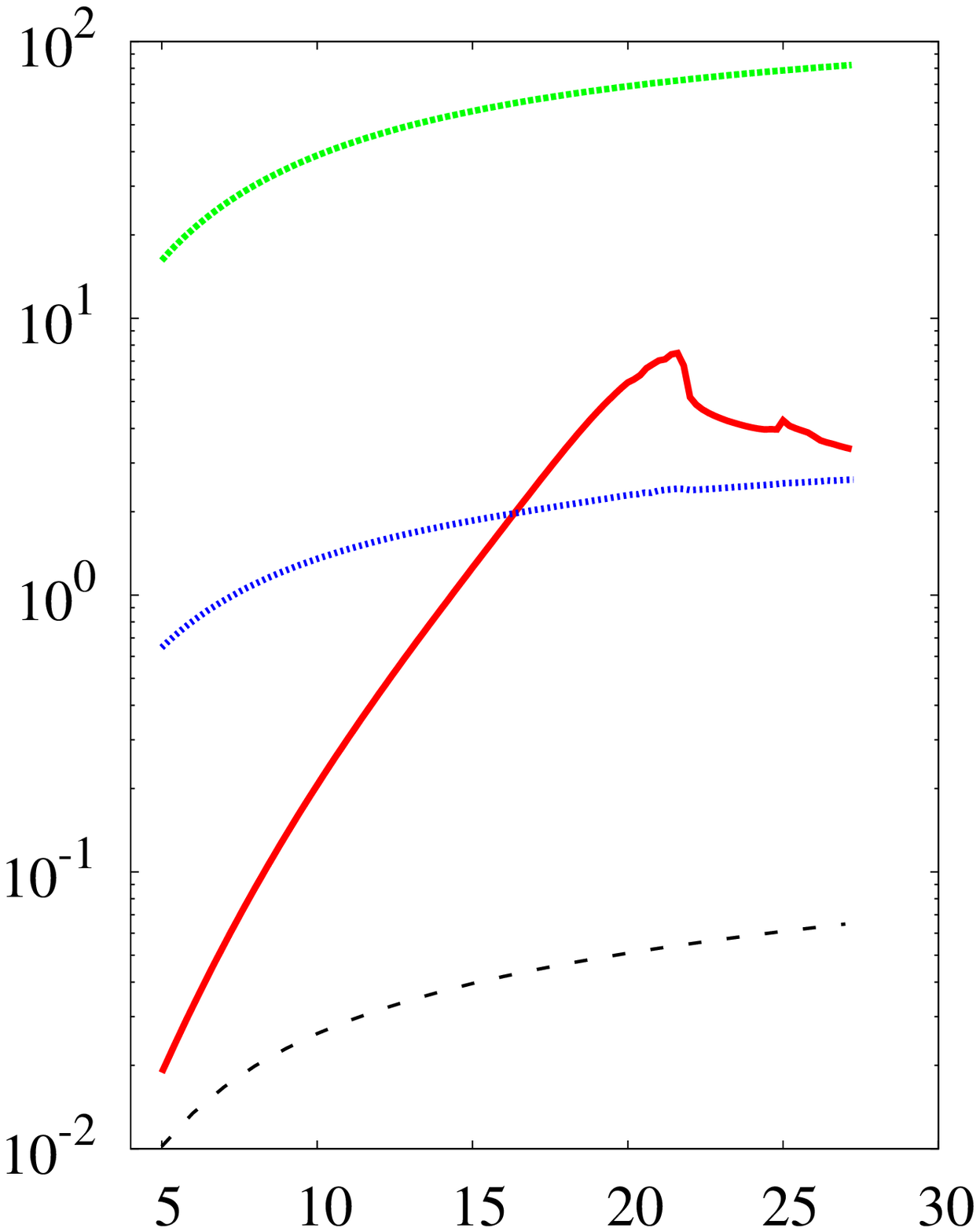,height=3.8in,width=3.2in}}}
\put(30,-15){\makebox(0,0)[bl]{\large{$\tan\beta$}}}
\put(40,220){\makebox(0,0)[bl]{\large{$\sigma(\tilde{t_1} \tilde{t}^\star_1)$ [pb]}}}
\put(-20,215){\makebox(0,0)[bl]{\large{$gg $}}}
\put(-26,150){\makebox(0,0)[bl]{\large{$\sum q\bar{q}$}}}
\put(0,80){\makebox(0,0)[bl]{\large{$b\bar{b}$}}}
\put(20,100){\makebox(0,0)[bl]{\large{$m_{A^0} = 370$~GeV}}}
\put(20,40){\makebox(0,0)[bl]{\large{no-Higgs}}}
\put(150,0){\mbox{\psfig{file=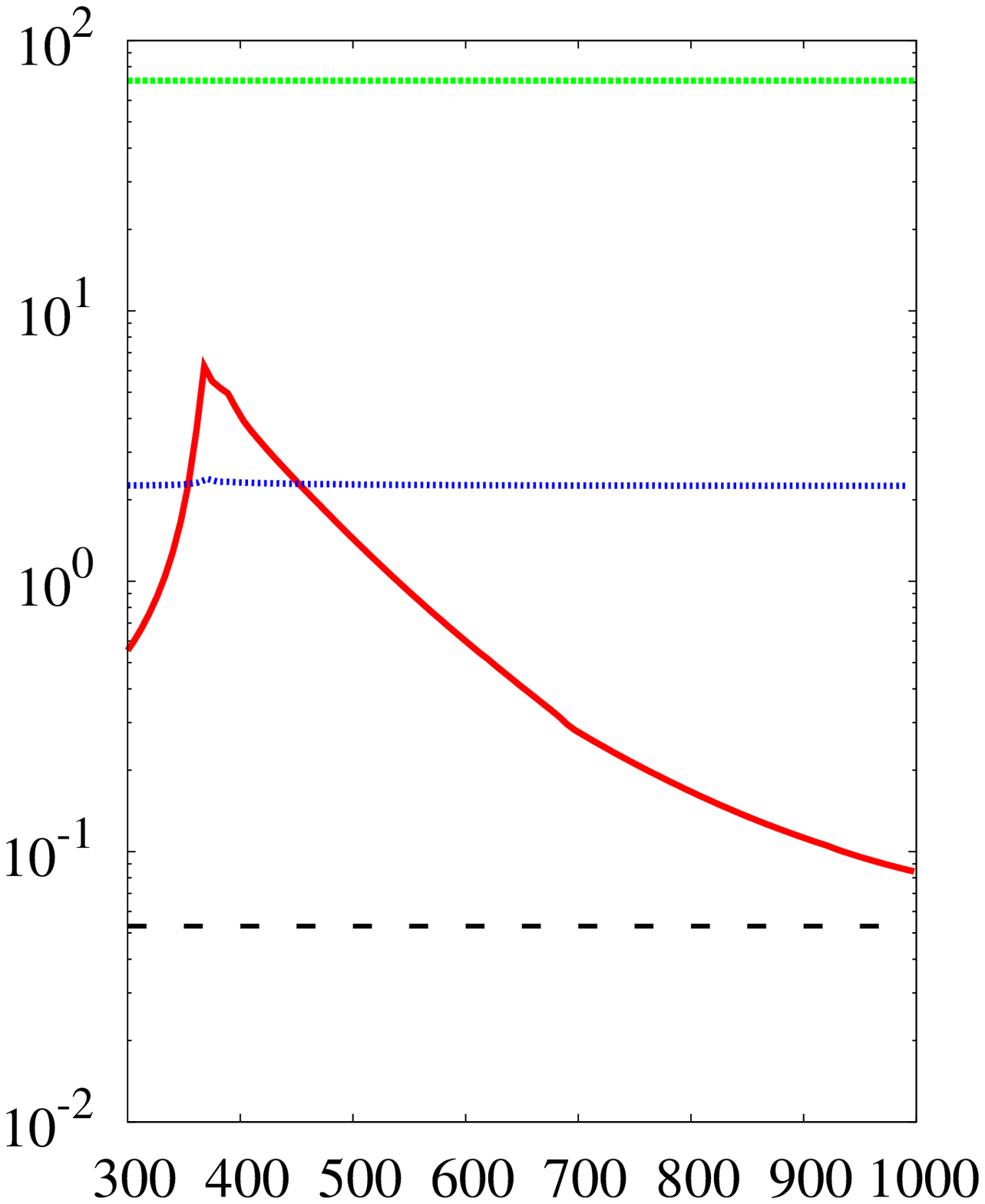,height=3.8in,width=3.2in}}}
\put(250,-15){\makebox(0,0)[bl]{\large{$m_{A^0}\,\,(GeV) $}}}
\put(250,215){\makebox(0,0)[bl]{\large{$\sigma(\tilde{t_1} \tilde{t}^\star_1)$ [pb]}}}
\put(250,240){\makebox(0,0)[bl]{\large{$gg $}}}
\put(280,120){\makebox(0,0)[bl]{\large{$b\bar{b} $}}}
\put(280,167){\makebox(0,0)[bl]{\large{$\sum q\bar{q}$}}}
\put(210,103){\makebox(0,10)[bl]{\large{$\tan\beta = 21$}}}
\put(220,50){\makebox(0,0)[bl]{\large{no-Higgs}}}
\end{picture}
\caption{ Production rates for $\tilde{t}_1 \tilde{t}^*_1$ pair 
production as   a function of (left) $\tan\beta$ and (right) $m_{A^0}$. 
{ Other SUSY parameters are chosen to be $M_{SUSY} = 350$
  GeV,} $M_2 = 200$ GeV, $m_{\tilde{g}} = 550$
  GeV, $\mu = -960$ GeV, $ A_0 = 750$ GeV.}
\label{fig-st1st1}
\end{figure}

We present the results of diagonal squark production of 
$\tilde{b}_1 \tilde{b}_1^*$ in Fig.~\ref{fig-sb1sb1} and 
$\tilde{t}_1 \tilde{t}_1^*$ in Fig.~\ref{fig-st1st1}, respectively.
At the LHC, the diagonal pair
production of $\widetilde{t}_1 \widetilde{t}_1^*$ and 
$\widetilde{b}_1 \widetilde{b}_1^*$ is dominated by $gg$ fusion.
 However, it is noted that
the $b\bar b$-initiated subprocess already surpass the light 
$q\bar q$-initiated channels when $\tan\beta \agt 17$, because of the
enhancement from the Higgs boson couplings to the bottom quark
as well as from the large coupling of Higgs to a pair of squarks.

The cross sections $\sigma(b\bar{b}, gg, \sum q\bar{q} \to 
\tilde{t}_1 \tilde{t}_1^*$ and $\tilde{b}_1 \tilde{b}_1^*$)
are plotted as a function of pseudoscalar Higgs 
mass $m_{A^0}$ for {a large $\tan\beta$ in the right panel and as a 
function of $\tan\beta$ at around the resonance 
$m_{A^0} \approx 2 m_{\tilde{q}}$, where $q = b,t$, in the left panel.}
The size of these cross sections depend strongly on the 
Higgs mass and $\tan\beta$.
We note that indeed the  leading order cross section 
$\sigma(b\bar{b}\to \tilde{t}_1 \tilde{t}_1^*$ and $\tilde{b}_1 \tilde{b}_1^*)$
increases like {$\tan^2\beta$} for large values of $\tan\beta$. 
To understand this we  
separate various contributions to the production rate. 
The bottom-induced Drell-Yan contribution proceeds through an 
s-channel photon or Z boson. The coupling to the photon is 
independent of $\tan\beta$ and the dependence of the Z couplings is
small for large enough values of the pseudoscalar Higgs mass.
A second set of s-channel production processes is made possible by the incoming
bottom quarks and their finite Yukawa couplings to 
CP-even Higgs bosons $h^0$, $H^0$, { but the }
CP-odd pseudoscalar exchange is forbidden by the CP 
symmetry of the final state. {The couplings 
$h^0\tilde{b}_1\tilde{b_1}$ and $H^0\tilde{b}_1\tilde{b_1}$
 depend strongly on $\tan\beta$ and on other MSSM parameters such as $\mu$ 
and soft trilinear terms $A_{t,b}$. Note that the dependence on $\tan\beta$
and $\mu$ parameter also comes in through $\Delta_b$ corrections.}

The relative size of the Higgs contribution compared to 
Drell-Yan process show up at large 
$\tan\beta$ and near the resonance. Note that the interference between Higgs 
contributions and the QCD are zero due to color structure.
The process $b\bar{b} \to \tilde{q}_1 \tilde{q}^*_1$ 
receives EW contributions from t-channel exchange of a neutralino
and gluino if q =b and chargino exchange contribution if q = t. 
The corresponding Feynman diagrams are shown in 
Fig.~\ref{fig:qqsts-diagrams}(c) and (d). 
Combining s-channel and t-channel including Higgs contributions, 
we see that they interfere constructively. 
Since QCD contributions dominate even after inclusion of the electroweak
diagrams, the overall behavior of the total cross sections does
not change much. For example, for $\tan\beta \sim 20$, the EW increase the 
cross section for the pair production of $\tilde{b}_1 \tilde{b}_1$ squarks
by about 2\,pb, whereas they only contribute 24\% to the total cross section.

\begin{figure}[htbp] 
\begin{picture}(320,260)
\put(-80,0){\mbox{\psfig{file=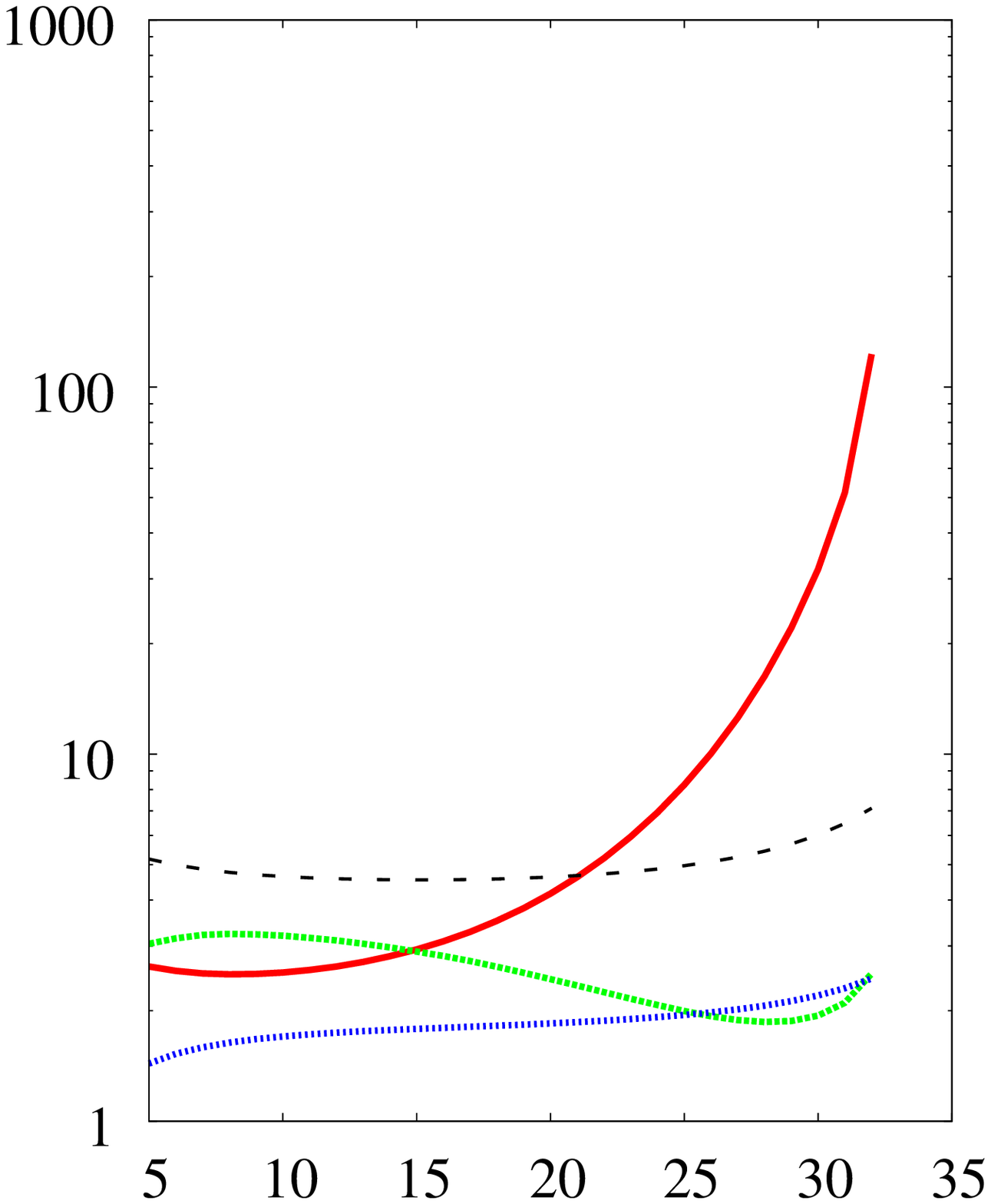,height=3.8in,width=3.2in}}}
\put(30,-15){\makebox(0,0)[bl]{\large{$\tan\beta$}}}
\put(-32,240){\makebox(0,0)[bl]{\large{$\sigma(\tilde{b}_1 \tilde{b}_2^* + \tilde{b}_2 \tilde{b}_1^* )$ in [fb]}}}
\put(91,165){\makebox(0,0)[bl]{\large{$b\bar{b}$}}}
\put(-25,33){\makebox(0,0)[bl]{\large{$\sum q\bar{q}$}}}
\put(55,52){\makebox(0,0)[bl]{\large{$gg$}}}
\put(-25,80){\makebox(0,10)[bl]{\large{no-Higgs}}}
\put(150,0){\mbox{\psfig{file=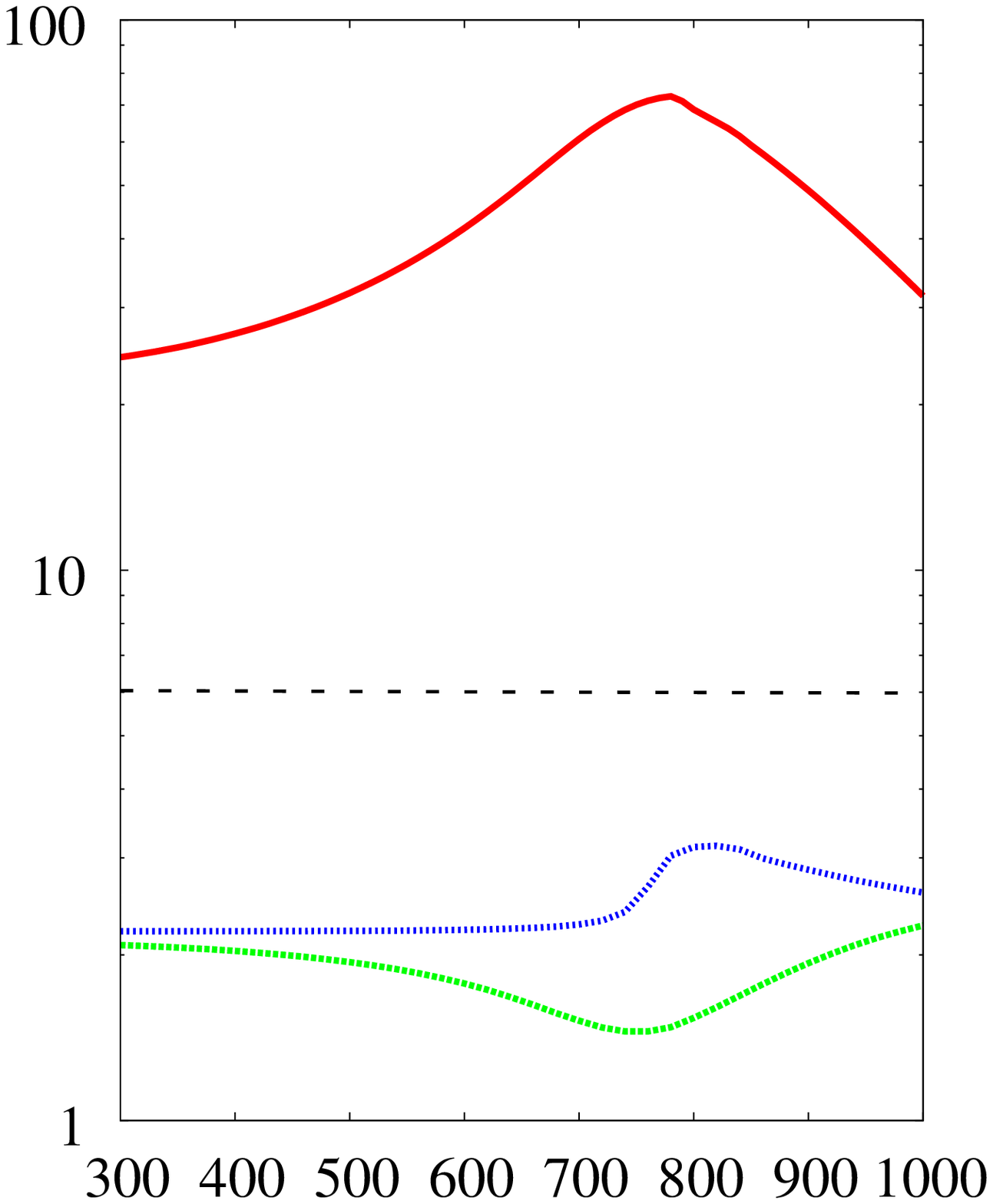,height=3.8in,width=3.2in}}}
\put(250,-15){\makebox(0,0)[bl]{\large{$m_{A^0}\,\,(GeV) $}}}
\put(200,160){\makebox(0,0)[bl]{\large{
$\sigma (\tilde{b}_1\tilde{b}_2^* + \tilde{b}_2\tilde{b}_1^*)$ in [fb]}}}
\put(230,205){\makebox(0,0)[bl]{\large{$b\bar{b}$}}}
\put(280,80){\makebox(0,0)[bl]{\large{$\sum q\bar{q}$}}}
\put(300,50){\makebox(0,0)[bl]{\large{$gg$}}}
\put(205,230){\makebox(0,0)[bl]{\large{$\tan\beta = 30$}}}
\put(205,120){\makebox(0,10)[bl]{\large{no-Higgs}}}
\end{picture}
\caption{ The cross sections of $pp\to\tilde{b}_1 \tilde{b}_2^* + 
\tilde{b}_2 \tilde{b}_1^*$ pair production as a function of (left) 
$\tan\beta$ and (right) $m_{A^0}$.  The SUSY parameters are chosen to 
be $M_{SUSY} = 400$ GeV, $M_2 = 200$ GeV, $m_{\tilde{g}} = 1000$ GeV, 
$\mu = -500$ GeV, $ A_0 = 500$ GeV.
\label{fig-nodiagonal1}
} 
\end{figure}

\begin{figure}[htbp] 
\begin{picture}(320,260)
\put(-80,0){\mbox{\psfig{file=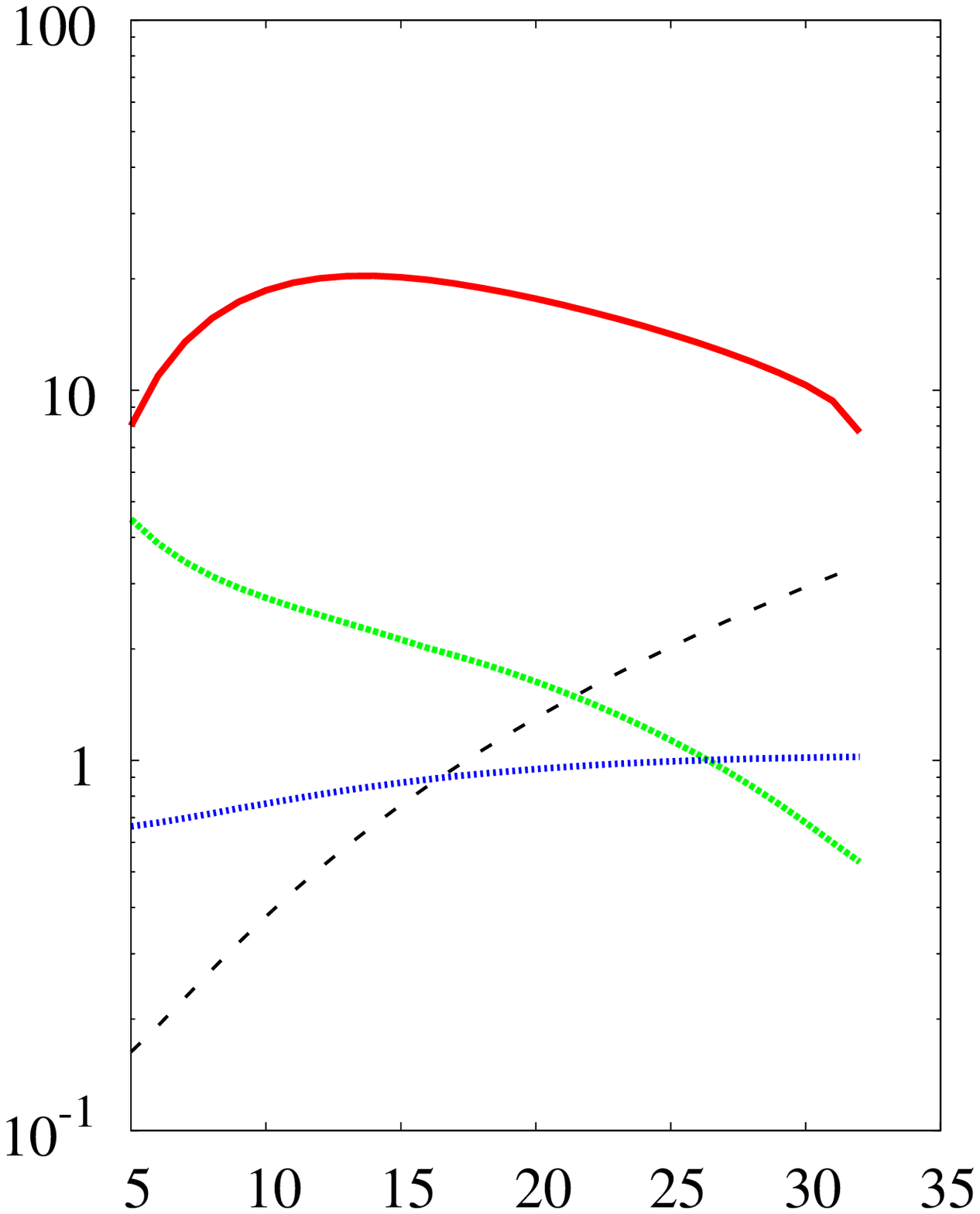,height=3.8in,width=3.2in}}}
\put(30,-15){\makebox(0,0)[bl]{\large{$\tan\beta$}}}
\put(-32,240){\makebox(0,0)[bl]{\large{$\sigma(\tilde{t}_1 \tilde{t}_2^* + \tilde{t}_2 \tilde{t}_1^* )$ in [fb]}}}
\put(91,165){\makebox(0,0)[bl]{\large{$b\bar{b}$}}}
\put(-25,95){\makebox(0,0)[bl]{\large{$\sum q\bar{q}$}}}
\put(5,132){\makebox(0,0)[bl]{\large{$gg$}}}
\put(-5,60){\makebox(0,10)[bl]{\large{no-Higgs}}}
\put(150,0){\mbox{\psfig{file=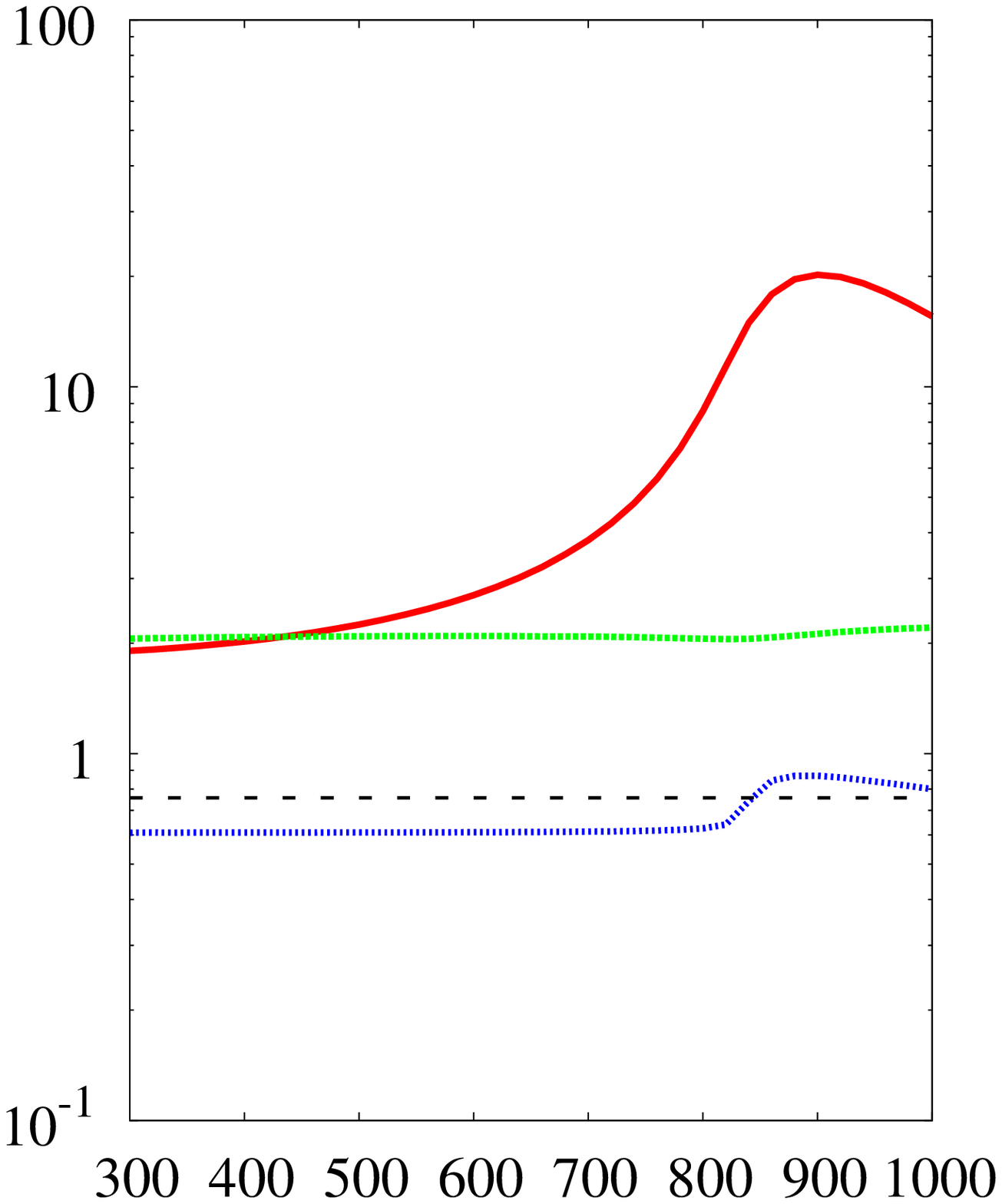,height=3.8in,width=3.2in}}}
\put(250,-15){\makebox(0,0)[bl]{\large{$m_{A^0}\,\,(GeV) $}}}
\put(200,235){\makebox(0,0)[bl]{\large{
$\sigma (\tilde{t}_1\tilde{t}_2^* + \tilde{t}_2\tilde{t}_1^*)$ in [fb]}}}
\put(270,150){\makebox(0,0)[bl]{\large{$b\bar{b}$}}}
\put(280,70){\makebox(0,0)[bl]{\large{$\sum q\bar{q}$}}}
\put(310,130){\makebox(0,0)[bl]{\large{$gg$}}}
\put(205,200){\makebox(0,0)[bl]{\large{$\tan\beta = 18$}}}
\put(205,98){\makebox(0,10)[bl]{\large{no-Higgs}}}
\end{picture}
\caption{ The cross sections of $pp\to\tilde{t}_1 \tilde{t}_2^* + 
\tilde{t}_2 \tilde{t}_1^*$ pair production as a function of (left) 
$\tan\beta$ and (right) $m_{A^0}$. The SUSY parameters are chosen to be 
$M_{SUSY} = 400$ GeV, $M_2 = 200$ GeV, $m_{\tilde{g}} = 1$ TeV, 
$\mu = -500$ GeV, $ A_0 = 500$ GeV.} 
\label{fig-nodiagonal2}
\end{figure}

In gluon-gluon fusion, the non-diagonal squark 
$\tilde{q}_i \tilde{q}^*_j$ with $i\neq j$ cannot be produced 
at the lowest order but via loop diagrams (see Fig.~2). 
The production cross section is 
therefore of the order ${\cal O}(\alpha^4_s+\alpha^2_s\alpha^2)$. 
This higher-order cross section 
will be small compared to $q\bar{q}$ tree level contribution. 
The $b\bar{b}$-fusion into non-diagonal squark pair
proceeds via CP-even Higgs bosons $h^0$, $H^0$, {
and} CP-odd Higgs boson 
$A^0$ in s-channel, gluino and neutralino in t-channel for 
$\tilde{b}_1 \tilde{b}^*_2$ and chargino in  t-channel for 
$\tilde{t}_1 \tilde{t}^*_2$ pair production. The cross sections 
$\sigma(\tilde{b}_1 \tilde{b}^*_2 + 
\tilde{b}_2 \tilde{b}^*_1)$  and 
$\sigma(\tilde{t}_1 \tilde{t}^*_2 + 
\tilde{t}_2 \tilde{t}^*_1)$ are plotted as a function 
of $\tan\beta$ in the left panel {
and as a function of $m_A$ in the right panel}
of Fig.~\ref{fig-nodiagonal1} and 
Fig.~\ref{fig-nodiagonal2},  respectively. 
As we can see, {the cross sections for both processes
are small due to the absence of tree-level $gg$ fusion and suppression from
the phase space.} For intermediate
values of $\tan\beta$ due to negative interference term between 
s-channel and t-channel diagrams 
all processes are comparable in size. 
For large $\tan\beta$, $\Delta_b$ effect induces a large enhancement for
$b\bar{b} \to \tilde{q}_1 \tilde{q}^*_2 + 
\tilde{q}_2 \tilde{q}^*_1 $ with $q = b,t$ by an order of
magnitude. 
The size of the $b\bar b$-initiated subprocess
is already above the $gg$ or $q\bar q$-initiated channels for 
$\tan\beta > 5$, as shown in Fig.~\ref{fig-nodiagonal1} {
and Fig.~\ref{fig-nodiagonal2}.}  
Note that in the case of 
$b\bar{b} \to \tilde{b}_1 \tilde{b}^*_2$, 
we can see a spectacular enhacement for large $\tan\beta$ (see 
Fig.~\ref{fig-nodiagonal1} left). { The origin of 
this enhancement is due to a factor of $\tan\beta$ in the $b\bar{b}A^0$ 
coupling and another factor of 
$\tan\beta$ in the $A^0\tilde{b}_1 \tilde{b}^*_2$ coupling,} which
results in $\tan^4\beta$ enhancement in cross section.
This is a very
interesting result that we cannot ignore the $b\bar b$-initiated 
subprocess with $s$-channel Higgs boson exchange.



\begin{table*}[h!]
\begin{centering}
\begin{ruledtabular}
\begin{tabular}{lc|cccc|cccc}
\multicolumn{2}{c|}{} & \multicolumn{4}{c|}{$\mu < 0$, $\quad$ $\Delta_b = -0.76$} & \multicolumn{4}{c}{$\mu > 0$, $\quad$ $\Delta_b = 0.66$} \tabularnewline
\multicolumn{2}{c|}{$\sigma$ {[}fb{]}} %
  & QCD     & EW    & Higgs   & Total  & QCD & EW    & Higgs   & Total   \\
\hline
  & $b\bar{b}\to \widetilde{b}_1\widetilde{b}^*_1$ & 11.7  & 16.6  & 410.2  & 437.10 & 0.37  & 1.24  & 0.007  & 1.68  \tabularnewline
%
%
& $b\bar{b}\to \widetilde{t}_1\widetilde{t}^*_1$ & 4.5 & 3.10 & 201.8 & 210.80 & 2.80 & 2.21 & 7.01 & 9.70 \tabularnewline
%
%
& $b\bar{b}\to \widetilde{b}_1\widetilde{b}^*_2 + \rm{h.c}$ & - & 2.65 & 110.7 &125.10 & - & 1.81 & 0.52 & 3.23  \tabularnewline
& $b\bar{b}\to \widetilde{t}_1\widetilde{t}^*_2 + \rm{h.c}$ &  - & 0.47 & 9.8 & 10.80 & - & 0.43 & 8.35  & 9.60  \tabularnewline
\hline\hline
& $c\bar{b}\to \widetilde{t}_1\widetilde{b}^*_1 $ & - & 0.76 $\times 10^{-3}$ & 27.85 & 27.86 & - & 0.94 $\times 10^{-4}$ & 1.71 $\times 10^{-2}$ & 1.72 $\times 10^{-2}$ \tabularnewline
& $u\bar{d}\to \widetilde{t}_1\widetilde{b}^*_1 $ & - & 10.16 & 45.95 & 56.21 & - & 1.77  & 6 $\times 10^{-2}$ & 1.84 \tabularnewline
%
& $c\bar{b}\to \widetilde{t}_1\widetilde{b}^*_2 + \rm{h.c} $ & - & 0.36 $\times 10^{-4}$ & 0.32 & 0.33 & - & 0.83 $\times 10^{-4}$ & 2.72 $\times 10^{-2}$ & 2.73 $\times 10^{-2}$ \tabularnewline
%
& $u\bar{d}\to \widetilde{t}_1\widetilde{b}^*_2 + \rm{h.c} $ & - & 4.8 $\times 10^{-2}$ & 1.71 & 1.72 & - & 0.25 $\times 10^{-2}$ & 0.87 & 0.88  \tabularnewline
\end{tabular}
\end{ruledtabular}
\par\end{centering}
\caption{The
effect of the sign of the $\mu$ parameter on the production cross sections (in fb). The SUSY parameters are chosen to be $A_0 = m_{\tilde{g}} = |\mu| = 1 $ TeV, $M_{SUSY} = 490 $ GeV and $M_2 = 200$ GeV, $\tan\beta = 20$ and the Higgs masses are taken at the resonance.\label{tab:separate}}
\end{table*}

Table~\ref{tab:separate} shows the cross sections including QCD, EW and Higgs effects. It is clear that Higgs effects enhance the cross sections for negative $\mu$ by one to two orders of magnitudes. While they reduce the cross sections for a positive $\mu$. Hence, we can see that the EW and QCD contributions are comparable for diagonal pair production

\begin{figure}[h!] 
\begin{picture}(320,260)
\put(-80,0){\mbox{\psfig{file=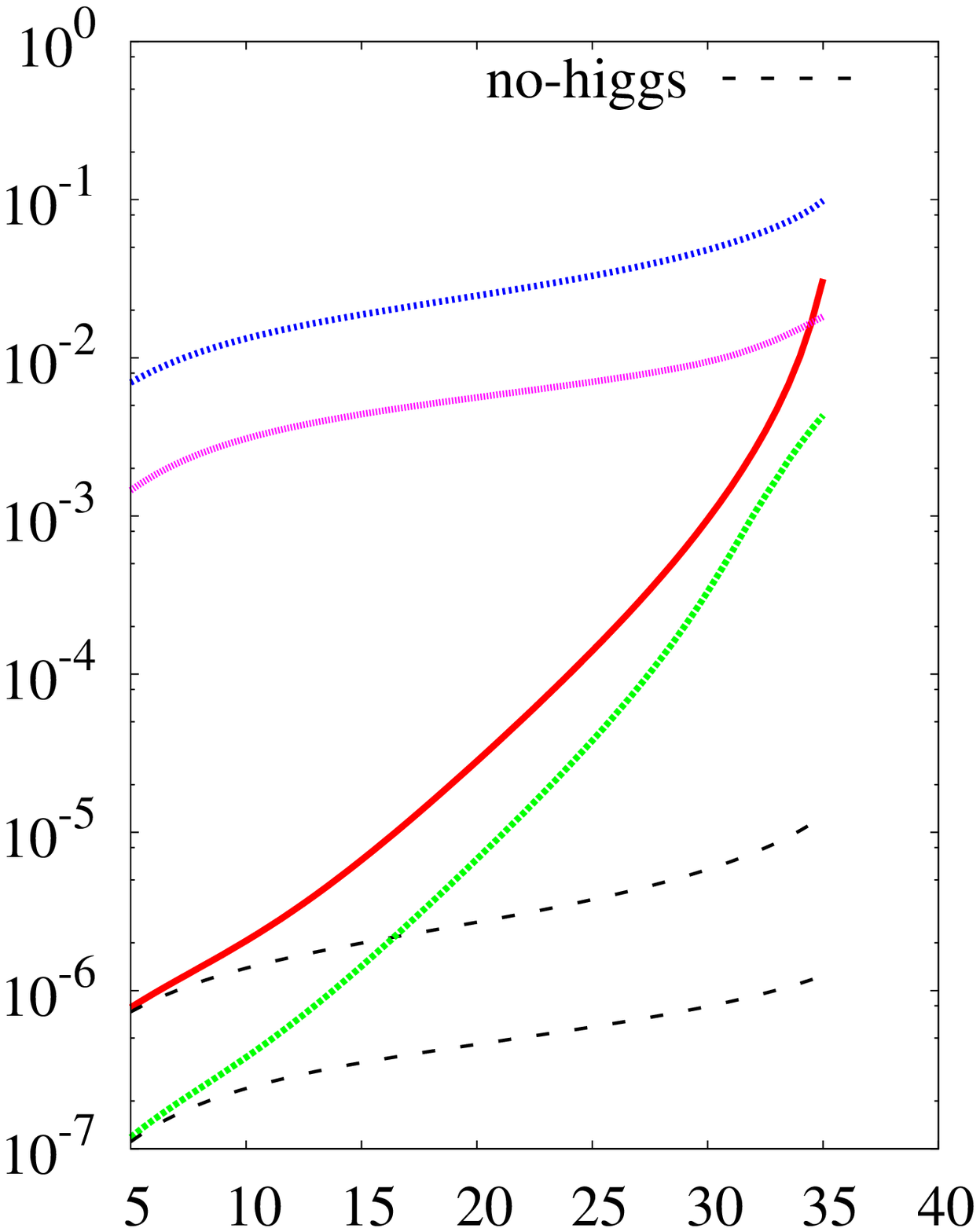,height=3.8in,width=3.2in}}}
\put(30,-15){\makebox(0,0)[bl]{\large{$\tan\beta$}}}
\put(-32,240){\makebox(0,0)[bl]{\large{$\sigma$ in [pb]}}}
\put(0,210){\makebox(0,0)[bl]{\large{$u\bar{d}\to\tilde{t}_1 \tilde{b}^*_1 $}}}
\put(0,169){\makebox(0,0)[bl]{\large{$u\bar{d}\to\tilde{t}_2 \tilde{b}^*_1 $}}}
\put(0,130){\makebox(0,0)[bl]{\large{$c\bar{b}\to\tilde{t}_1 \tilde{b}^*_1 $}}}
\put(85,130){\makebox(0,0)[bl]{\large{$\tilde{t}_2 \tilde{b}^*_1 $}}}
%
\put(150,0){\mbox{\psfig{file=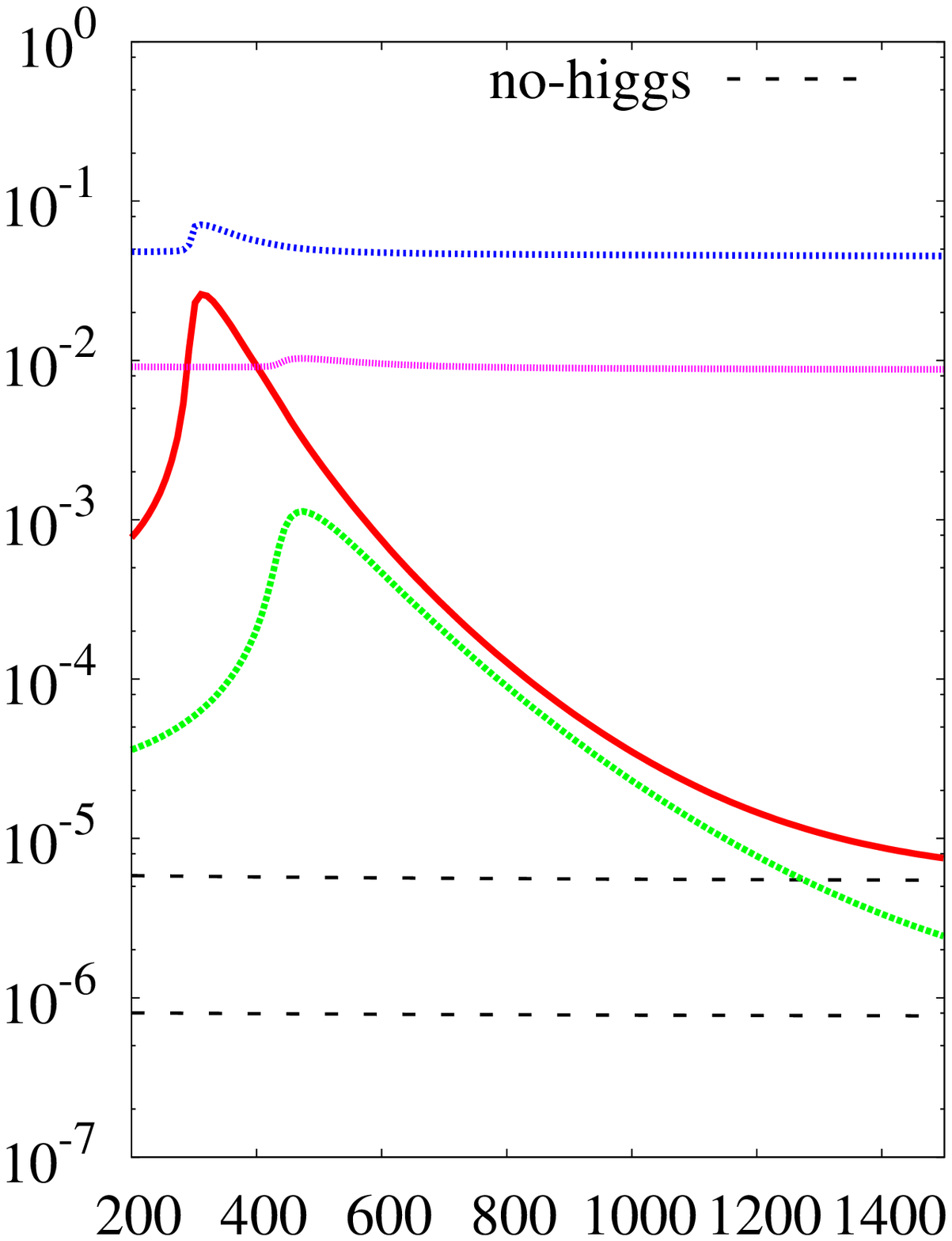,height=3.8in,width=3.2in}}}
\put(250,-15){\makebox(0,0)[bl]{\large{$m_{H^\pm}\,\,(GeV) $}}}
\put(200,240){\makebox(0,0)[bl]{\large{$\sigma$ in [pb]}}}
\put(260,215){\makebox(0,0)[bl]{\large{$u\bar{d}\to\tilde{t}_1 \tilde{b}^*_1 $}}}
\put(260,191){\makebox(0,0)[bl]{\large{$u\bar{d}\to\tilde{t}_2 \tilde{b}^*_1 $}}}
\put(240,160){\makebox(0,0)[bl]{\large{$c\bar{b}\to\tilde{t}_1 \tilde{b}^*_1 $}}}
\put(215,110){\makebox(0,0)[bl]{\large{$c\bar{b}\to\tilde{t}_2 \tilde{b}^*_1 $}}}
\put(225,85){\makebox(0,0)[bl]{\large{$c\bar{b}\to\tilde{t}_1 \tilde{b}^*_1 $}}}
\put(225,55){\makebox(0,0)[bl]{\large{$c\bar{b}\to\tilde{t}_2 \tilde{b}^*_1 $}}}
\put(225,25){\makebox(0,0)[bl]{\large{$\tan\beta = 30$}}}
\end{picture}
\caption{ The cross sections of mixed squark $pp\to\tilde{t}_i
  \tilde{b}^*_j$ pairs production as a function of (left) $\tan\beta$
  and (right) $m_{H^\pm}$. The SUSY parameters are
  chosen to be $M_{SUSY} = 200$ GeV, $M_2 = 200$ GeV, $m_{\tilde{g}} =
  1000$ GeV, $\mu = -200$ GeV, $ A_0 = 200$ GeV. {Note that for non-diagonal production $q\bar{q'}\to \tilde{t}_i \tilde{b}^*_j$ we have taken into account their hermetic conjugate}.}

\label{fig-stisbj}
\end{figure}

Stop-sbottom pair production is dominated by the $W$ exchange diagram
with $u\bar d$-initiated  subprocess. The charged-Higgs contribution 
through $c\bar b$-initiated  subprocess can get to comparable size
of cross section at very large $\tan\beta \agt 35$, as 
shown in Fig.~\ref{fig-stisbj}.  
Note that the charged-Higgs couples to
the $c \bar b$ quarks with an enhancement from $\tan\beta$ but also 
with a suppression from $V_{cb}$. {As we can see,
large $\tan\beta$ limit overcome easily the $V_{cb}$ suppression}.


\section{Conclusions}

We have reviewed the theoretical status of {
squark pair production} at the LHC. The evaluation of the full electroweak 
contributions has been described in details. The Higgs effect enhance
the cross section by about 10-25\%, which is comparable to the size
of NLO correction as well as to the PDF uncertainties.
In the case of non-diagonal squarks production, 
we have seen some enhancement 
for large $\tan\beta$. Those processes can be used to extract some
information on the squarks mixing angles.
One concludes that LO electroweak contribution has to be taken into account
for any reliable prediction.

\section*{Acknowledgments} 
A.A is supported by the National Science of Theoretical Studies-Taipei 
under contract \# 980528731. 
R.B is supported by National Cheng Kung University Grant No. HUA 97-03-02-063.
R.B acknowledges the KEK theory exchange program for physicists in Taiwan and the very kind hospitality at KEK.
K.C. was supported in parts by the NSC
under Grant Nos. (96-2628-M-007-002-MY3), 
by the NCTS, by the Boost Program of the NTHU, and
by WCU program through the NRF funded by the MEST (R31-2008-000-10057-0).
T.C. was supported by the NSC under Grant no. 98-2112-M-001-014-MY3.


\appendix

\section{Feynman Rules}
\label{appendix-hsc}
In this appendix, we give the relevant Feynman rules for our study.
\subsection{Higgs-squark couplings}

In this appendix we collect the couplings of the $h^0$, $H^0$ and $A^0$ 
bosons to the squarks $\tilde{q}_i$ with q= t, b and i= 1,2, 
which are relevant for our analysis. by using the mixing matrix 
which rotates the left- and 
right-handed squark fields $\tilde{q}_L$ and $\tilde{q}_R$, 
into mass eigenstates $\tilde{q}_i$ as
 Squark couplings to Higgs boson  are given by
where $\alpha$ is the mixing angle in the neutral Higgs sector, we have used $c_\alpha = \cos\alpha$, $s_\alpha = \sin\alpha$,  $c_{\alpha+\beta} = \cos(\alpha+\beta)$, $s_{\alpha+\beta} = \sin(\alpha+\beta)$, $s_\beta = \sin\beta$ and $c_\beta = \cos\beta$ as abbreviations. We note the following properties of the above couplings  $g_{h^0 \tilde{q}_1 \tilde{q}_2 } = g_{h^0 \tilde{q}_2 \tilde{q}_1 }$, $g_{H^0 \tilde{q}_1 \tilde{q}_2 } = g_{H^0 \tilde{q}_2 \tilde{q}_1 }$ and $g_{A^0 \tilde{q}_1 \tilde{q}_2 } =  - g_{A^0 \tilde{q}_2 \tilde{q}_1 }$.  Squark couplings to quarks and either charginos $\tilde\chi^{\pm}$ or neutralinos $\tilde\chi^{0}$ are straightforward, but somewhat more complicated by the mixing angles associated with the $\tilde\chi^{\pm}$ and $\tilde\chi^{0}$ mass eigenstates.\\

\subsubsection{Squark -- Squark -- $h^0$}
\begin{eqnarray}
\hat G_1 = \left(\! \begin{array}{cc}
  \frac{g\,m_Z^{}}{{\rm c}_W^{}}\,C_{qL}\,s_{\alpha+\beta} 
                        -\sqrt{2}\;m_q\,Y^q  
\left\{\begin{array}{c}c_\alpha \\ -s_\alpha\end{array}\right\}
  & -\frac{1}{\sqrt{2}}\,Y^q\,\big( A_q\left\{\begin{array}{c}c_\alpha \\
-s_\alpha\end{array}\right\}  
                                +\mu \left \{\begin{array}{c}
s_\alpha \\ c_\alpha \end{array} \right\}\big) 
  \\[4mm]
  -\frac{1}{\sqrt{2}}\,Y^q\,\big( A_q \left\{\begin{array}{c} c_\alpha \\
-s_\alpha \end{array} \right\} 
                              +\mu \left \{\begin{array}{c}
s_\alpha \\ c_\alpha \end{array} \right\}\big) 
  & \frac{g\,m_Z^{}}{{\rm c}_W^{}}\,C_{qR}\,s_{\alpha+\beta} 
                         -\sqrt{2}\,m_q\,Y^q \left\{\begin{array}{c} c_\alpha \\ -s_\alpha \end{array} \right\}
\end{array} \right)
\label{eq:GLR1}
\end{eqnarray}

for {\small $\Big\{\!\begin{array}{c} \mbox{\footnotesize up} \\[-1mm]
\mbox{\footnotesize down} \end{array} \!\Big\}$}
type squarks respectively. $\alpha$ is the mixing angle in the CP even
neutral Higgs boson sector.
$Y^q$ are the Yukawa couplings:   
\begin{equation}
  Y^t = \frac{g\,m_t}{\sqrt{2}\:m_{W}\sin\beta}, \hspace{8mm} 
  Y^b = \frac{g\,m_b}{\sqrt{2}\:m_{W}\cos\beta}.
\end{equation}
\subsubsection{Squark -- Squark -- $H^0$}

\begin{eqnarray}
\hat G_2 = \left(\! \begin{array}{cc}
  -\frac{g\,m_Z^{}}{{\rm c}_W^{}}\,C_{qL}\,c_{\alpha+\beta} 
                        -\sqrt{2}\;m_q\,Y^q  
\left\{\begin{array}{c}s_\alpha \\ c_\alpha\end{array}\right\}
  & -\frac{1}{\sqrt{2}}\,Y^q\,\big( A^q\left\{\begin{array}{c}s_\alpha \\
c_\alpha\end{array}\right\}  
                                -\mu \left \{\begin{array}{c}
c_\alpha \\ s_\alpha \end{array} \right\}\big) 
  \\[4mm]
  -\frac{1}{\sqrt{2}}\,Y^q\,\big( A_q \left\{\begin{array}{c} s_\alpha \\
c_\alpha \end{array} \right\} 
                              -\mu \left \{\begin{array}{c}
c_\alpha \\ s_\alpha \end{array} \right\}\big) 
  & -\frac{g\,m_Z^{}}{{\rm c}_W^{}}\,C_{qR}\,s_{\alpha+\beta} 
                         -\sqrt{2}\,m_q\,Y^q \left\{\begin{array}{c} s_\alpha \\ c_\alpha \end{array} \right\}
\end{array} \right)
\label{eq:GLR2}
\end{eqnarray}

Notice that $G_2$ can be obtained from $G_1$ by the 
replacement $\alpha \to\alpha+\pi/2$, i.e ${\rm c}_\alpha \to {\rm s}_\alpha$ and 
${\rm s}_\alpha \to -{\rm c}_\alpha$.

\subsubsection{Squark -- Squark -- $A^0$}

\begin{eqnarray}
\hat{G}_3=i\frac{gm_q}{2m_W} \left(\begin{array}{cc} 0
& -A_q\left\{\begin{array}{c}\cot\beta \\ \tan\beta
\end{array}\right\} -\mu \\ A_q\left\{\begin{array}{c}\cot\beta
\\ \tan\beta \end{array}\right\} +\mu & 0 \end{array} \right)
\end{eqnarray}

\subsubsection{Squark -- Squark -- $G^0$}
\begin{eqnarray}
\hat{G}_4=i\frac{gm_q}{2m_W} \left(\begin{array}{cc} 0
& -A_q +\mu\left\{\begin{array}{c}\cot\beta \\ \tan\beta
\end{array}\right\} \\ A_q -\mu\left\{\begin{array}{c}\cot\beta
\\ \tan\beta \end{array}\right\} & 0 \end{array} \right)
\end{eqnarray}

\subsubsection{ squark -- squark -- $H^\pm$}
\begin{eqnarray}
\hat{G}^{\tilde{t}}_5=(\hat{G}^{\tilde{b}}_5)^T =
\frac{g}{\sqrt{2}m_W}\left(\begin{array}{cc}
m_b^2\tan\beta +m_t^2\cot\beta - m^2_W s_{2\beta}  &  m_t(A_t\cot\beta +\mu) \\
m_b(A_b\tan\beta +\mu) & 2m_tm_b/\sin2\beta
\end{array} \right)
\end{eqnarray}

\subsubsection{ squark -- squark -- $G^\pm$}
\begin{eqnarray}
\hat{G}^{\tilde{t}}_6=(\hat{G}^{\tilde{b}}_6)^T =
\frac{g}{\sqrt{2}m_W}\left(\begin{array}{cc}
m_t^2 -m_b^2 - m^2_W c_{2\beta}& m_t(A_t -\mu\cot\beta) \\
m_b(\mu\tan\beta -A_b) & 0
\end{array} \right)
\end{eqnarray}

\subsection{quark-Squark-neutralino and quark-Squark-chargino}
\subsubsection{quark -- squark -- neutralino}
\begin{eqnarray}
{\cal{A}}_{im}^{\tilde{q}}&=& -\frac{1}{3\sqrt{2}c_W s_\beta s_W }
\left\{\begin{array}{c}
-4m_W s_\beta R^{\tilde{u}}_{i2} N^*_{m1} + 3c_W m_u N^*_{m4}R^{\tilde{u}}_{i1}
\\ 2 c_\beta M_W s_W N^*_{m1} R^{\tilde{d}}_{i2} + 3c_W m_d  N^*_{m3}R^{\tilde{d}}_{i1}
\end{array} \right\},\\ 
{\cal{B}}_{im}^{\tilde{q}}&=& -\frac{1}{3\sqrt{2}c_W s_\beta s_W}
\left\{\begin{array}{c}
3 c_W m_u N_{m4} R_{i2}^{\tilde{u}} + m_W s_\beta R_{i1}^{\tilde{u}}(s_W N_{m1} + 3c_W N_{m2})
\\ 3c_W m_d R_{i2}^{\tilde{d}} N_{m3} + c_\beta m_W R_{i1}^{\tilde{d}}(s_W N_{m1} - 3c_W N_{m2})
\end{array} \right\}
\end{eqnarray}
Here $N$ is the $4\times 4$ unitary matrix diagonalizing the
neutral gaugino-higgsino mass matrix~\cite{Haber:1984rc}.
\subsubsection{quark -- squark -- chargino}
\begin{eqnarray}
{\cal{L}}_{ik}^{\tilde{q}}= U^*_{k2} U^*_{i1}\left\{\begin{array}{c}
\frac{m_d}{\sqrt{2}c_\beta m_W}  \\ 
\frac{m_u}{\sqrt{2}s_\beta m_W}
\end{array} \right\},\quad 
{\cal{K}}_{ik}^{\tilde{q}}=-\left\{\begin{array}{c}
\frac{1}{2m_W s_\beta} (2m_W s_\beta R^{\tilde u}_{i1} V_{k1} - \sqrt{2} m_u R^{\tilde u}_{i2} V_{k2})
 \\ 
\frac{1}{2m_W c_\beta} (2m_W c_\beta R^{\tilde d}_{i1} U_{k1} - \sqrt{2} m_d R^{\tilde d}_{i2} U_{k2})
\end{array} \right\}.
\end{eqnarray}
Here $U$ and $V$ are the $2\times 2$ unitary matrices
diagonalizing the charged gaugino--higgsino mass matrix~\cite{Haber:1984rc}.

\section{Production rates}

\label{cross-sections}


The production of squark pairs, as initiated by $b\bar{b}$ annihilation, involves gluon, photon, Z, $W$ and Higgs bosons in the s-channel as well as gluino and neutralino exchanges in the t-channel. 
Since left-and right-squarks generally have different masses we present the differential cross section for each subprocess separately in the mass basis. 
The spin and color averages are taken into account.
i,j = 1...2, m,n = 1...4 and k,l =1...2.
\subsection{$b\,\overline{b}\rightarrow\widetilde{b}_i\, \widetilde{b}^*_j$}
The differential cross section for 
$b\,\overline{b}\rightarrow\widetilde{b}_i\, \widetilde{b}^*_i$ (diagonal)
is given by
\begin{eqnarray} 
\frac{d\hat\sigma(b\,\overline{b}\rightarrow\widetilde{b}_i\,
\widetilde{b}^{\star}_i)}{d\hat t}&=&\frac{2\pi}{9\hat s^2}
\left(\hat t\hat u-m_{\widetilde{b}_i}^4 \right) \\\nonumber
&\times&\bigg[ \frac{2 \alpha_s^2}{\hat s^2}+  \frac{9 \alpha^2}{\hat s^2}
e^4_b + \bigg( \frac{4\alpha \alpha_s e^2_b}{\hat s(\hat t-m^2_{\widetilde{g}})}-
\frac{2\alpha^2_s}{3 \hat s (\hat t-m^2_{\widetilde{g}})}\bigg) \big[({\cal{R}}^{\widetilde{b}}_{i1})^2 + ({\cal{R}}^{\widetilde{b}}_{i2})^2\big] \\\nonumber
&+&\frac{\big[({\cal{A}}^{\widetilde{b}}_{ik})^2+({\cal{B}}^{\widetilde{b}}_{ik})^2\big]}{(\hat
  t-m^2_{\widetilde{\chi}^0_m})} 
\bigg(\frac{3\alpha^2 e^2_b}{2\hat s} + \frac{2\alpha \alpha_s}{\hat s}\bigg)
+\frac{3\alpha^2 e_b}{2} \frac{(\hat s-M^2_Z)}{\hat s} D^2_Z g_{Z \widetilde{q}_i \widetilde{q}_i}
\\\nonumber
&-& \frac{3 \alpha^2 e_b }{2 \hat s} \frac{(\hat u-\hat t)}{\left(\hat t\hat
    u-m_{\widetilde{b}_i}^4 \right)} \bigg((\hat s-M^2_h)g_{h bb}D_h
(G_1)_{ii} + 
(\hat s-M^2_H)g_{Hbb} D_H (G_2)_{ii}\bigg)
\\\nonumber
&+& \frac{9 \alpha^2}{4} (C^2_{qL} + C^2_{qR}) g^{2}_{Z \widetilde{b}_i \widetilde{b}_i} D^2_Z
+ \frac{2 \alpha\alpha_s}{3 } \big(C_{qL}({\cal{R}}^{\widetilde{b}}_{i1})^2 + C_{qR}({\cal{R}}^{\widetilde{b}}_{i2})^2\big) g_{Z \widetilde{b}_i \widetilde{b}_i} D^2_Z
\frac{(\hat s-M^2_Z)}{(\hat t-m^2_{\widetilde{g}})}
\\\nonumber &+& 
\frac{9 \alpha^2}{4\left(\hat t\hat u-m_{\widetilde{b}_i}^4 \right)} 
 \bigg|g_{hbb} D_{h} (G_1)_{ii} + g_{Hbb} D_{H} (G_{2})_{ii}\bigg|^2
\\\nonumber
&+& \frac{\alpha^2_s}{(\hat t-m^2_{\widetilde{g}})^2}
\bigg[ ({\cal{R}}^{\widetilde{q}}_{i1})^4 + ({\cal{R}}^{\widetilde{q}}_{i2})^4  + 
\frac{2 m^2_{\widetilde{g}} \hat s}{\left(\hat t\hat u-m_{\widetilde{b}_i}^4 \right)}({\cal{R}}^{\widetilde{q}}_{i1})^2({\cal{R}}^{\widetilde{q}}_{i2})^2
\bigg]\\\nonumber
&+& \frac{9 \alpha^2}{8 (\hat t-m^2_{\widetilde{\chi}^0_k})(\hat t-m^2_{\widetilde{\chi}^0_m})}
\bigg[ ({\cal{A}}^{\widetilde{q}}_{ik})^4 + ({\cal{B}}^{\widetilde{q}}_{im})^4  + 
\frac{2 m_{\widetilde{\chi}^0_k} m_{\widetilde{\chi}^0_m}  \hat s}{\left(\hat t\hat u-m_{\widetilde{b}_i}^4\right)}({\cal{A}}^{\widetilde{q}}_{ik})^2({\cal{B}}^{\widetilde{q}}_{im})^2\bigg]\\\nonumber
&+& \bigg[\frac{ \alpha\alpha_s }{3} \frac{\hat s m_{\widetilde{g}} }{(\hat t
  -m^2_{\widetilde{g}})}({\cal{R}}^{\widetilde{q}}_{i1})({\cal{R}}^{\widetilde{q}}_{i2}) -\frac{4\alpha^2 }{3}g_{Z \widetilde{q}_i \widetilde{q}_i}(C_{qL}+C_{qR})(\hat s-M^2_Z)(\hat u-\hat t)D_Z 
\bigg]
 \\\nonumber&\times&
\frac{1}{\left(\hat t\hat u-m_{\widetilde{b}_i}^4 \right)}\bigg[ (\hat s-m^2_h)
g_{hbb} D_{h} (G_1)_{ii} +  (\hat s-m^2_H) g_{Hbb} D_{H} (G_{2})_{ii}\bigg]\\\nonumber
\\\nonumber
&-& \frac{\alpha^2 g_{Z \widetilde{q}_i \widetilde{q}_i}}{4
}\frac{\big[({\cal{A}}^{\widetilde{q}}_{ik})^2
  C_{bL}+({\cal{B}}^{\widetilde{q}}_{ik})^2
  C_{bR}\big]}{(\hat t-m^2_{\widetilde{\chi}^0_k})} 
(\hat s-M^2_{Z})D^2_{Z}\\\nonumber
&+&\frac{3 \alpha^2 }{4 (\hat t-m^2_{\widetilde{\chi}^0_k})}
{\cal{A}}^{\widetilde{q}}_{ik}{\cal{B}}^{\widetilde{q}}_{ik}
m_{\widetilde{\chi}^0_k} \hat s \bigg[ (\hat s-M^2_{h^0})g_{hbb} D^2_{h} (G_1)_{ii}
+ (\hat s-M^2_{H^0}) g_{Hbb} D^2_{H} (G_2)_{ii}\bigg] 
\bigg ]
\end{eqnarray}
Where $D_{\Phi}^{-1}=\hat{s}-m_{\Phi}^2+i m_{\Phi} \Gamma_{\Phi}$.
The imaginary
part in the $D_{\Phi}^{-1}$ is the Breit-Wigner 
prescription for regulating the $\Phi$ pole.

%

while the differential cross section for off-diagonal ($i \!\not =j$) is 
given by
\begin{eqnarray} 
\frac{d\hat\sigma(b\,\overline{b}\rightarrow\widetilde{b}_i\,
\widetilde{b}^{\star}_j)}{d\hat t}&=&\frac{2\pi}{9\hat s^2}
\left(\hat t\hat u-m_{\widetilde{q}_i}^2 m_{\widetilde{q}_j}^2\right)\bigg[
\frac{9 \alpha^2}{4 } (C^2_{qL} + C^2_{qR}) g^{2}_{Z \widetilde{q}_i \widetilde{q}_j} D^2_Z
\\\nonumber
&+& \frac{2 \alpha\alpha_s}{3 } \big(C_{qL}({\cal{R}}^{\widetilde{q}}_{i1})^2 + C_{qR}({\cal{R}}^{\widetilde{q}}_{j2})^2\big) 
g_{Z \widetilde{q}_i \widetilde{q}_j} D^2_Z
\frac{(\hat s-M^2_Z)}{(\hat t-m^2_{\widetilde{g}})}
\\\nonumber &+& \frac{1}{\left(\hat t\hat u-m_{\widetilde{q}_i}^2 m_{\widetilde{q}_j}^2 \right)} 
\frac{9 \alpha^2 }{16} \big|g_{hbb} D_{h} (G_1)_{ij} + g_{Hbb} D_{H} (G_{2})_{ij}\big|^2\\\nonumber
&+& \frac{\alpha^2_s}{(\hat t-m^2_{\widetilde{g}})^2}
\bigg[ ({\cal{R}}^{\widetilde{q}}_{i1})^4 + ({\cal{R}}^{\widetilde{q}}_{j2})^4  + 
\frac{2 m^2_{\widetilde{g}} \hat s}{\left(\hat t\hat u-m_{\widetilde{q}_i}^2 m_{\widetilde{q}_j}^2\right)}({\cal{R}}^{\widetilde{q}}_{i1})^2({\cal{R}}^{\widetilde{q}}_{j2})^2
\bigg]\\\nonumber
&+& \frac{9 \alpha^2}{8 (\hat t-m^2_{\widetilde{\chi}^0_k})(\hat t-m^2_{\widetilde{\chi}^0_m})}
\bigg[ ({\cal{A}}^{\widetilde{q}}_{ik})^4 + ({\cal{B}}^{\widetilde{q}}_{jm})^4  + 
\frac{2 m_{\widetilde{\chi}^0_k} m_{\widetilde{\chi}^0_m} \hat s}{\left(\hat t\hat u-m_{\widetilde{q}_i}^2 m_{\widetilde{q}_j}^2\right)}({\cal{A}}^{\widetilde{q}}_{ik})^2({\cal{B}}^{\widetilde{q}}_{jm})^2
\bigg]\\\nonumber
&+& \bigg[\frac{ \alpha\alpha_s}{3} \frac{\hat s m_{\widetilde{g}} }{(\hat t
  -m^2_{\widetilde{g}})}({\cal{R}}^{\widetilde{q}}_{i1})({\cal{R}}^{\widetilde{q}}_{j2}) -\frac{4\alpha^2 }{3}g_{Z \widetilde{q}_i \widetilde{q}_j}(C_{qL}+C_{qR})(\hat s-M^2_Z)(\hat u-\hat t)D_Z 
\bigg]
 \\\nonumber&\times&
\frac{1}{\left(\hat t\hat u-m_{\widetilde{q}_i}^2 m_{\widetilde{q}_j}^2
  \right)}\bigg[ (\hat s-m^2_h) g_{hbb} D_{h} (G_1)_{ij} +  (\hat s-m^2_H)g_{Hbb} D_{H} (G_{2})_{ij}\bigg]\\\nonumber
\\\nonumber
&-& \frac{\alpha^2 g_{Z \widetilde{q}_i
    \widetilde{q}_j}}{4}\frac{\big[({\cal{A}}^{\widetilde{q}}_{ik})^2
  C_{bL}+({\cal{B}}^{\widetilde{q}}_{jk})^2
  C_{bR}\big]}{(\hat t-m^2_{\widetilde{\chi}^0_k})} (\hat s-M^2_{Z})D^2_{Z}\\\nonumber
&+&\frac{3 \alpha^2}{4 (\hat t-m^2_{\widetilde{\chi}^0_k})}
{\cal{A}}^{\widetilde{q}}_{ik}{\cal{B}}^{\widetilde{q}}_{jk}
m_{\widetilde{\chi}^0_k} \hat s \bigg[ (\hat s-M^2_{h^0}) g_{hbb} D^2_{h}
(G_1)_{ij} + 
(\hat s-M^2_{H^0}) g_{Hbb} D^2_{H} (G_2)_{ij}\bigg]
\end{eqnarray}
where summations over $i, j$.

\subsubsection{$b\,\overline{b}\rightarrow\widetilde{t}_i\, \widetilde{t}^*_j$ }
Similarly, the differential cross sections  for 
$\widetilde{t}_i \widetilde{t}_i^*$ and $\widetilde{t}_i \widetilde{t}_j^*$ 
are respectively given by
\begin{eqnarray}
\label{eq11}
\frac{d\hat\sigma(b\,\overline{b}\rightarrow\widetilde t_i\,
\widetilde{t}^{\star}_i)}{d \hat t}&=&\frac{2\pi}{9 \hat s^2}
\left( \hat t \hat u-m_{\widetilde{t}_i}^4 \right)
\bigg[\frac{2 \alpha^2_s}{s^2} + \frac{e^2_t \alpha^2}{\hat s^2} + \frac{\alpha^2}{8} (C^2_{bL} + C^2_{bR})g^2_{Z \widetilde{t}_i \widetilde{t}_i} D^2_Z \\\nonumber
&+&\frac{9\,\hat s \alpha^2}{4(\hat t \hat u-m_{\widetilde{t}_i}^4)} \bigg| g_{h \widetilde{t}_i \widetilde{t}_i} g_{hbb} D_h + g_{H \widetilde{t}_i \widetilde{t}_i} g_{Hbb} D_H\bigg|^2\\\nonumber 
&+& \frac{9 \alpha^2}{8 (\hat t-m^2_{\widetilde{\chi}^+_k})(t-m^2_{\widetilde{\chi}^+_l})}
\bigg( ({\cal{L}}^{\widetilde{t}}_{ik})^4 + ({\cal{K}}^{\widetilde{t}}_{il})^4  + 
\frac{2 m_{\widetilde{\chi}^+_k} m_{\widetilde{\chi}^+_l}  \hat s}{( \hat t
  \hat u-m_{\widetilde{t}_i}^4 )}({\cal{L}}^{\widetilde{t}}_{ik})^2({\cal{K}}^{\widetilde{t}}_{il})^2
\bigg)\\\nonumber
&+&\frac{2\alpha \alpha_s}{\hat s} \frac{\big[({\cal{L}}^{\widetilde{t}}_{ik})^2 +
  ({\cal{K}}^{\widetilde{t}}_{il})^2\big]}{(\hat t-m^2_{\widetilde{\chi}^+_k})}
+\frac{\alpha^2 e_t}{6 \hat s} (C_{bL}+C_{bR})g_{Z \widetilde{t}_i \widetilde{t}_i}
(\hat s-M^2_Z) D^2_Z
\\\nonumber
&-& 3 \alpha^2 \bigg[\frac{\hat s}{2}
\frac{({\cal{L}}^{\widetilde{t}}_{ik}{\cal{K}}^{\widetilde{t}}_{ik})
  m_{\widetilde{\chi}^+_k}}{(\hat t-m^2_{\widetilde{\chi}^+_k})} 
-  \frac{(\hat t-\hat u)}{( \hat t \hat u-m_{\widetilde{t}_i}^4 )}
\bigg(\frac{ e_t}{\hat s}
-\frac{3}{2} g_{Z \widetilde{t}_i \widetilde{t}_i} (C_{bL} + C_{bR}) (\hat s-M^2_Z) D^2_Z\bigg)
\bigg]\\\nonumber 
&\times&\bigg( g_{h \widetilde{t}_i \widetilde{t}_i}  g_{hbb} (\hat s-M^2_{h})
D^2_{h} + g_{H \widetilde{t}_i \widetilde{t}_i}  g_{Hbb} (\hat s-M^2_{H}) D^2_{H}\bigg)
\\\nonumber
&-&\frac{3\alpha^2}{2} \frac{(\hat s-M^2_Z)}{(\hat t-m^2_{\widetilde{\chi}^+_k})} \big[(C_{bL}({\cal{K}}^{\widetilde{t}}_{ik})^2 + C_{bR} ({\cal{L}}^{\widetilde{t}}_{ik})^2\big] D^2_Z
-\frac{\alpha^2}{2 \hat s} \frac{[({\cal{K}}^{\widetilde{t}}_{ik})^2 +
  ({\cal{L}}^{\widetilde{t}}_{ik})^2]}{(\hat t-m^2_{\widetilde{\chi}^+_k})} 
\bigg]
\end{eqnarray}

\begin{eqnarray}
\label{eq12}
\frac{d\hat\sigma(b\,\overline{b}\rightarrow\widetilde t_i\,
\widetilde{t}^{\star}_j)}{d \hat t}&=&\frac{2\pi}{9 \hat s^2}
\left( \hat t \hat u-m_{\widetilde{t}_i}^2 m_{\widetilde{t}_j}^2\right)
\bigg[
\frac{\alpha^2}{8} (C^2_{bL} + C^2_{bR})g^2_{Z \widetilde{t}_i \widetilde{t}_j} D^2_Z \\\nonumber
&+&\frac{9\,\hat s \alpha^2}{4(\hat t \hat u-m_{\widetilde{t}_i}^2 m_{\widetilde{t}_j}^2)} \bigg| g_{h \widetilde{t}_i \widetilde{t}_j} g_{hbb} D_h + g_{H \widetilde{t}_i \widetilde{t}_j} g_{Hbb} D_H + g_{A \widetilde{t}_i \widetilde{t}_j} g_{Abb} D_A \bigg|^2\\\nonumber 
&+& \frac{9 \alpha^2}{8 (\hat t-m^2_{\widetilde{\chi}^+_k})(\hat t-m^2_{\widetilde{\chi}^+_l})}
\bigg( ({\cal{L}}^{\widetilde{t}}_{ik})^4 + ({\cal{K}}^{\widetilde{t}}_{jl})^4  + 
\frac{2 m_{\widetilde{\chi}^+_k} m_{\widetilde{\chi}^+_l}  \hat s}{( \hat t \hat u-m_{\widetilde{t}_i}^2 m_{\widetilde{t}_j}^2)}({\cal{L}}^{\widetilde{t}}_{ik})^2({\cal{K}}^{\widetilde{t}}_{jl})^2
\bigg)\\\nonumber
&-& 3 \alpha^2 \bigg[\frac{\hat s}{2}
\frac{({\cal{L}}^{\widetilde{t}}_{ik}{\cal{K}}^{\widetilde{t}}_{jk})
  m_{\widetilde{\chi}^+_k}}{(\hat t-m^2_{\widetilde{\chi}^+_k})} 
+ \frac{3}{2}\frac{(\hat t-\hat u)}{( \hat t \hat u-m_{\widetilde{t}_i}^2 m_{\widetilde{t}_j}^2 )}
\bigg( g_{Z \widetilde{t}_i \widetilde{t}_j} (C_{bL} + C_{bR}) 
(\hat s-M^2_Z) D^2_Z\bigg)
\bigg]\\\nonumber 
&\times&\bigg( g_{h \widetilde{t}_i \widetilde{t}_j}  g_{hbb} (\hat s-M^2_{h})
D^2_{h} + g_{H \widetilde{t}_i \widetilde{t}_j}  g_{Hbb} (\hat s-M^2_{H}) D^2_{H} +
g_{A \widetilde{t}_i \widetilde{t}_j}  g_{Abb} (\hat s-M^2_{A}) D^2_{A}\bigg)
\\\nonumber
&-&\frac{3\alpha^2}{2} \frac{(\hat s-M^2_Z)}{(\hat t-m^2_{\widetilde{\chi}^+_k})} \big[(C_{bL}({\cal{K}}^{\widetilde{t}}_{jk})^2 + C_{bR} ({\cal{L}}^{\widetilde{t}}_{ik})^2\big] D^2_Z
\bigg]
\end{eqnarray}
\subsection{stop-sbottom production}
\begin{eqnarray}
\label{eq16}
\frac{d\hat\sigma(q\,\overline{q^\prime}\rightarrow\widetilde t_i\,
\widetilde{b}^{\star}_j)}{d \hat t}\nonumber &=&\frac{2\pi}{9 \hat s^2} |V_{qq^\prime}|^2 |V_{\widetilde{t_i} \widetilde{b}_j}|^2
\bigg[\frac{18 \alpha^2}{s^4_W}\left( \hat t u-m_{\widetilde{t}_i}^2 m_{\widetilde{b}_j}^2\right) D^2_W ({\cal{R}}^{\widetilde{b}}_{i1} {\cal{R}}^{\widetilde{t}}_{j1})^2 \\ &+& \frac{9 \alpha^2 m^2_b}{8 s^2_W} (G_5)^2_{ij} \hat{s} D^2_{H^\pm}  
- \frac{9 \alpha^2 m^2_b}{4s^2_w} (\hat t- \hat u) (G_5)_{ij} D_{H^\pm} D_{W^\pm}{\cal{R}}^{\widetilde{b}}_{i1} {\cal{R}}^{\widetilde{t}}_{j1}
\bigg] 
\end{eqnarray}
\subsection{$gg \to \widetilde{q}_i \widetilde{q}_i$}
\begin{eqnarray}
\label{eq17}
\frac{d\hat\sigma}{d\hat t}(gg\to\widetilde{q}_i\widetilde{q}^*_i) &=& \frac{\pi\alpha^2_s}{12 \hat s^2}
\Bigg[
\frac{2(m^4_{\tilde{q}}-\hat t \hat u)}{(m^2_{\tilde{q}}-\hat t)(m^2_{\tilde{q}}-\hat u)} + \frac{9[4\hat s(4m^2_{\tilde{q}} - \hat s) + (\hat u - \hat t)^2]}{8 \hat s^2} + \frac{7}{4}\nonumber \\
&-& \frac{(4m^2_{\tilde{q}}-\hat s)^2}{16(m^2_{\tilde{q}}-\hat t)(m^2_{\tilde{q}}-\hat u)} - \frac{7(4m^2_{\tilde{q}} + 4 \hat t - \hat s)}{32(\hat t -m^2_{\tilde{q}})} - \frac{7(4m^2_{\tilde{q}} + 4 \hat u - \hat s)}{32(\hat u -m^2_{\tilde{q}})}\nonumber\\
&+& \frac{9[(\hat t - \hat u)(4m^2_{\tilde{q}} + 4\hat t -\hat s) - 2(m^2_{\tilde{q}}-\hat u)(6m^2_{\tilde{q}} + 2\hat t - s)]}{32 \hat s (m^2_{\tilde{q}} -\hat t)}
\nonumber\\
&-& \frac{9[(\hat t - \hat u)(4m^2_{\tilde{q}} + 4\hat u -\hat s) + 2(m^2_{\tilde{q}}-\hat t)(6m^2_{\tilde{q}} + 2\hat u - \hat s)]}{32 \hat s (m^2_{\tilde{q}} -\hat u)}
\Bigg] 
\end{eqnarray}

\end{document}